\documentclass[fleqn,usenatbib]{mnras}

\usepackage{newtxtext,newtxmath}

\usepackage[T1]{fontenc}
\usepackage{ae,aecompl}

\usepackage{graphicx}	
\usepackage{amsmath}	
\usepackage{amssymb}	
\usepackage{bm}
\usepackage{times}
\usepackage[normalem]{ulem}

\newcommand{\hMpc}{\ensuremath{~h^{-1}\mathrm{Mpc}}}

\renewcommand{\arcmin}{\ensuremath{~\mathrm{arcmin}}}
\newcommand{\degSq}{\ensuremath{~\mathrm{deg}^2}}
\newcommand{\map}[1]{\ensuremath{#1 \times #1 \degSq}}

\newcommand{\nucut}{\ensuremath{\nu_\mathrm{cut}}}

\defcitealias{Takahashi2017}{T17}
\defcitealias{Matilla2017}{Z16}

\title[Self similar WL peaks]{The self similarity of weak lensing peaks}

\author[Davies et al.]{
Christopher T. Davies,$^{1}$\thanks{E-mail: christopher.t.davies@durham.ac.uk (CTD)}
Marius Cautun,$^{1}$
Baojiu Li$^{1}$
\\
$^{1}$Institute for Computational Cosmology, Department of Physics, Durham University, South Road, Durham DH1 3LE, UK\\
}

\pubyear{2019}

\begin{document}
\label{firstpage}
\pagerange{\pageref{firstpage}--\pageref{lastpage}}
\maketitle

\begin{abstract}
We study the statistics of weak lensing convergence peaks, such as their abundance and two-point correlation function (2PCF), for a wide range of cosmological parameters $\Omega_m$ and $\sigma_8$ within the standard $\Lambda$CDM paradigm, focusing on intermediate-height peaks with signal-to-noise ratio (SNR) of $1.5$ to $3.5$. 
We find that the cosmology dependence of the peak abundance can be described by a one-parameter fitting formula that is accurate to within $\sim3\%$. The peak 2PCFs are shown to feature a self-similar behaviour: if the peak separation is rescaled by the mean inter-peak distance, catalogues with different minimum peak SNR values have identical clustering, which suggests that the peak abundance and clustering are closely interconnected. A simple fitting model for the rescaled 2PCF is given, which together with the peak abundance model above can predict peak 2PCFs with an accuracy better than $\sim5\%$.
The abundance and 2PCFs for intermediate peaks have very different dependencies on $\Omega_m$ and $\sigma_8$, implying that their combination can be used to break the degeneracy between these two parameters.
\end{abstract}




\section{Introduction}
\label{sect:intro}

Gravitational lensing, the deflection of photon trajectory due to the presence of massive objects between the source and the observer, is sensitive to the total matter distribution in the Universe \citep{Bartelmann2001}. The total matter content can be expressed as the sum of baryons ($\Omega_b$) and dark matter ($\Omega_{\rm{DM}}$). Combined, these two components make up $\Omega_m\approx 30\%$ of the Universe's energy budget with an abundance ratio of $\Omega_{\rm{DM}} / \Omega_b \approx 5$ \citep{Planck2018}. While dark matter is the dominant matter component, it is also the more challenging one to detect since it is not directly observable. Gravitational lensing offers a promising probe which allows us to examine the distribution of dark matter in the Universe. 

In this paper we are interested in weak lensing (WL), which is the regime of gravitational lensing where the amplitude of light deflections is small. These deflections, caused by inhomogeneities of the total matter distribution, lead to distortions of the images of distant sources, such as background galaxies or the cosmic microwave background. While the distortions are small, they can be extracted with a careful statistical analysis of their correlations \citep[e.g.,][]{Bacon2000, Kaiser2000, VanWaerbeke2000, Wittman2000}. Nowadays, WL is widely used as a probe of the large-scale structures (LSS) of the Universe \citep{Albrecht2006,LSST2012,Amendola2013,Weinberg2013}.

While WL is only the broad name of a physical phenomenon, a plethora of statistics can be extracted from observations and cosmological simulations. For example, the cosmic shear (the distortion of the shapes of a lensed image) and convergence (the magnification of the magnitude of the lensed source), are usually quantified using their two-point statistics, though higher-order statistics contain extra information about the nonlinear LSS evolution. Analyses performed on the low-redshift universe using shear-shear correlations have led to strong cosmological parameter constraints \citep[e.g.,][]{Schneider2002,Semboloni2006,Hoekstra2006,Fu2008,Heymans2012,Kilbinger2013,Hildebrandt2017}. Lensing signatures are also present in the cosmic microwave background for which the lensing potential power spectra can be used as a cosmological test \citep[e.g.,][]{Planck2018}. WL has also been detected around cosmological objects such as galaxy clusters, which can be used to measure the cluster mass \citep[e.g.,][]{Gruen2014,Umetsu2014,Applegate2014,vonderLinden2014,Hoekstra2015,Tudorica2017}, and voids, for which void WL profiles can be obtained \citep[e.g.,][]{Melchior2014,Clampitt2015,Sanchez2017}.

Weak lensing can be used to test different cosmological models. For example, the convergence power spectrum, shear-shear correlations as well as WL by voids have been shown to be promising probes for constraining the dark energy equation of state or modified gravity models \citep[e.g.][]{Schmidt2008,Tsujikawa2008,Huterer2010,Cai2015,Barreira2015,Barreira2017,Baker2018,Cautun2018,vanUitert2018,Cautun2019}. While WL offers an independant measurement of the absolute cluster mass, this cannot be done for most clusters; one use of WL is to calibrate some observable-cluster-mass scaling relation, which is then used to infer cluster masses and hence the abundance of massive haloes, which is a powerful cosmological probe \citep[e.g.][]{Mantz2015,Umetsu2016,Pizzuti2017,Bocquet2018}. In addition, statistics such as WL Minkowski functionals and WL bispectrum have been used to constrain the sum of neutrino masses \citep[e.g.][]{Marques2018,Coulton2018}.

The most commonly used WL statistics are the shear correlation function and convergence power spectrum. These two-point statistics alone, however, cannot account for the non-Gaussian features introduced by the nonlinear evolution of structures in the Universe, and other statistics can provide additional and complementary information. 
In this paper we study one such additional probe, WL peaks, which are the maxima of the WL convergence field. The WL peak abundance is a good example of a statistic that contains complimentary information to two-point statistics \citep{Jain2000,Pen2003,Dietrich2010,Shirasaki2018}, and can be used to constrain cosmological parameters within $\Lambda$CDM \citep{Shan2012,VanWaerbeke2013,Shan2014,X.Liu2015}, to test alternative cosmological models such as modified gravity \citep{Cardone2013,X.Liu2016,Higuchi2016,Shirasaki:2016twn,Peel2018}, dark energy \citep{Giocoli2018}, and to measure the neutrino mass \citep{Li2018}. WL peaks can also be extracted from CMB lensing to provide cosmological constraints \citep{J.Liu2016b}. Various models have been developed to accurately describe high signal-to-noise-ratio (SNR) WL peaks \citep[e.g.,][]{Hamana2004, Hennawi2005, Maturi2005, Fan2010, Marian2012, Hamana2012, J.Liu2016, Shan2018, Wei2018}. 

In contrast to high WL peaks, there have been relatively few studies on the abundance of low and intermediate peaks \citep[see, e.g.,][]{Yang2011,Lin2015, Shirasaki2017}, which nevertheless contains rich cosmological information 
\citep{Dietrich2010, Kratochvil2010, Yang2011}, and even fewer on the spatial correlation of such peaks \citep[e.g.,][]{Marian2013,Shan2014}. Upcoming wide and deep field galaxy surveys such as {\sc euclid} \citep{Refregier2010} and {\sc lsst} \citep{LSST2009} will produce large high-resolution WL maps, with significant improvements compared to the current generation of WL observations. Understanding how WL peak statistics behave will be important if we want to maximise the cosmological information that can be gained from the new surveys. In particular, the higher source number density of these surveys will lead to a reliable determination of peak abundance and clustering down to low SNR values, so it is important to have accurate models to describe the statistics of low- and intermediate-height peaks.

In this work we study properties of WL peak statistics in $\Lambda$CDM, by modelling the peak abundance, peak two-point correlation functions (2PCFs) and the convergence rms fluctuation (convergence map standard deviation). Most importantly, we identify a universal self-similar behaviour in the peak 2PCF, which holds for a large range of peak heights and different cosmologies. 
The self-similarity is observed when expressing the 2PCF in terms of the angular separation divided by the mean peak separation, with the resulting rescaled 2PCFs lying on top of each other. We propose a general model that describes the abundance and clustering of WL peaks and that allows us to access cosmological information contained on non-linear scales. 

The structure of the paper is outlined as follows: in Section \ref{section: Theory, simulations and analysis pipeline} we briefly summarise the relevant theory for WL, describe the numerical simulations used to construct the WL maps, and introduce the statistics we use to study WL peaks. Next, in Section \ref{section: Weak lensing statistics} we present the WL peak abundance, WL peak 2PCF, and identify a self similarity in the peak 2PCF for a given fiducial cosmology. Then, in Section \ref{section: cosmo dependence} we give general fitting functions that describe the convergence rms fluctuation, peak abundance, 2PCF and its self similarity in $\Lambda$CDM for a large range of $\Omega_m$ and $\sigma_8$ values. We then show in Section \ref{section: reconstruction pipeline} that our model can accurately reproduce the original peak 2PCF. Finally, in Section \ref{section:GSN} we show that the self similarity of the 2PCF is robust to the inclusion of galaxy shape noise.

\section{Theory, simulations and analysis pipeline}
\label{section: Theory, simulations and analysis pipeline}

In this section we briefly summarise the essential elements of WL theory, and describe the simulations and methodology used in this work to study WL peak statistics.

\subsection{Theory}
The deflection of photon trajectories due to the mass of objects in the lens plane can be quantified using the lensing convergence $\kappa$,
\begin{equation}
    \centering
    \kappa = \frac{1}{2} \nabla^2 \Psi \, ,
    \label{Eq:convergence}
\end{equation}
where $\Psi$ is the 2D lensing potential given by:
\begin{equation}
    \Psi(\bm{\theta}) = \frac{D_{ls}}{D_l D_s} \frac{2}{c^2} \int_0^{z_s} \Phi(D_l\bm{\theta},z) dz \, .
    \label{Eq:lensing potential}
\end{equation}
In the above equation,
$\bm{\theta}$ is the observed angular position of the lensed image (in the Born approximation); $D_{s}$,$D_l$ and $D_{ls}$ are the angular diameter distances between the observer and source, observer and lens, and lens and source; $z_s$ is the source redshift; and $c$ is the speed of light. The symbol $\Phi$ denotes the 3D gravitational potential given by the Poisson equation,
\begin{equation}
    \nabla^2\Phi = 4 \pi G a^2 \delta \rho_m \, ,
    \label{Eq:Poisson equation}
\end{equation}
where $\delta\rho_m = \rho_m - \overline{\rho}_m$ with $\rho_m$ and $\overline{\rho}_m$ respectively the local and the background matter density;
$a$ is the scale factor; and $G$ is the gravitational constant. Eq. \eqref{Eq:convergence} therefore behaves as a projected version of the Poisson equation weighted by a geometric factor determined by the distances between the lens, source and observer.

The above equations describe the WL effect of a single lens. However, as the light travels between the source and the observer, it experiences gravitational lensing from the entirety of the mass distribution along its path.
The single lens equation can be generalised to multiple lenses as
\begin{equation}
    \kappa(\bm{\theta}) = \int_0^{z_s} W(z) \delta \rho_m(D_l(z)\bm{\theta},z) dz \, ,
    \label{general lensing with kernel}
\end{equation}
where $W(z)$ is the lensing kernel that accounts for the redshift distribution of the lenses. This kernel is given by
\begin{equation}
    W(z) = \frac{3 H_0^2 \Omega_m}{2 c} \frac{1+ z}{H(z)} \chi(z) \int_{z}^{z_s} \frac{dn}{dz_s} dz_s \frac{\chi(z_s) - \chi(z)}{\chi(z_s)} \, ,
\end{equation}
where $H(z)$ is the Hubble parameter and $H_0$ is its present-day value; $\Omega_m$ is the fractional total matter density at present-day; $\chi$ is the comoving distance; and $\frac{dn}{dz_s}$ is the redshift distribution of sources. 

\begin{figure*}
    \centering
    \includegraphics[width=\textwidth]{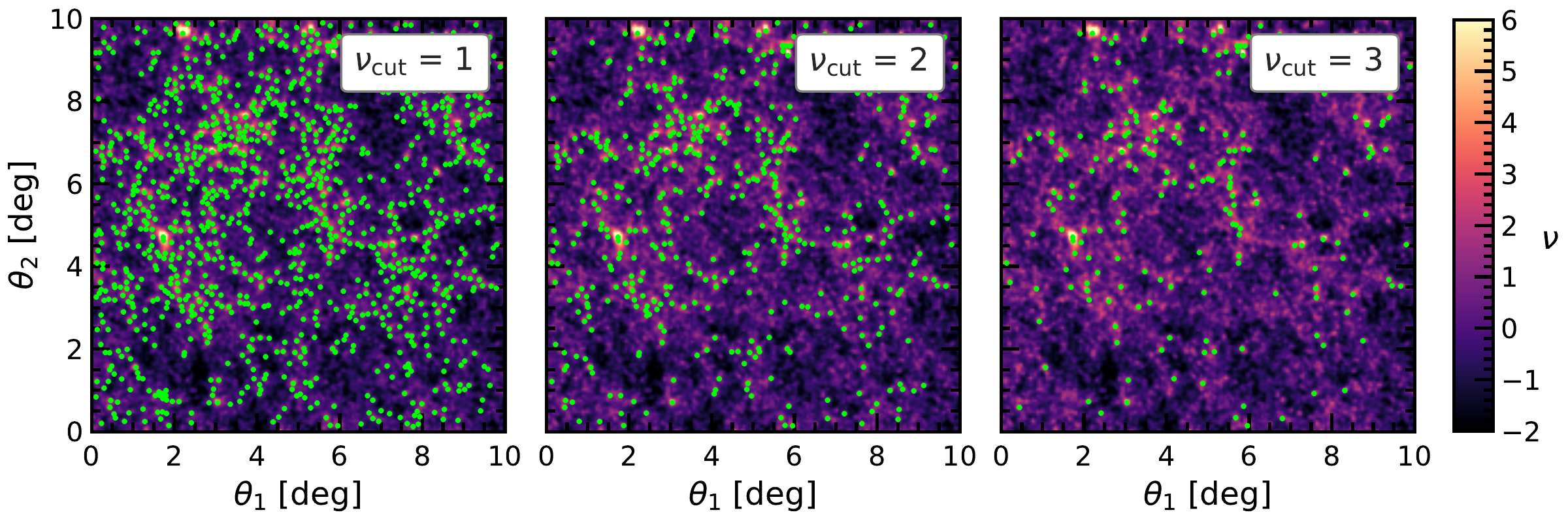}
    \vskip -.3cm
    \caption{An illustration of a WL convergence map and the distribution of peaks of different heights (green points). The convergence field is expressed in terms of the SNR, $\nu$, as indicated by the colour-bar on the right. Peaks with a height below $\nucut = 1, 2$ and $3$ are removed to produce the three peak catalogue shown in the three panels (all plotted on the same convergence map). The axes $\theta_1$ and $\theta_2$ give angular coordinates of the map in two orthogonal directions. The map is smoothed with $\theta_s = 2$ arcmin before the peaks are identified. We use 2 arcmin smoothing here for visualisation purposes, to clearly show the presence of WL peaks in a convergence map. As 1 arcmin smoothing (which is used in the main analysis of the paper) results in a large number of peaks that would reduce the clarity of the figure.}
    \vskip -.1cm
    \label{fig:peaks and kappa map}
\end{figure*}

As can be seen from Eq. \eqref{general lensing with kernel}, the WL convergence corresponds to the projected mass density contrast weighted by a geometric factor, and thus positive and negative $\kappa$ values correspond to overdense and underdense lines of sight. For self consistency across different convergence maps we define the SNR, $\nu$, as
\begin{equation}
    \nu = \frac{\kappa - \mu}{\sigma}, 
    \label{Eq:nu}
\end{equation}
where $\mu$ is the mean value of the convergence field of a given map, and $\sigma$ is its rms fluctuation. We have subtracted the map mean $\mu$ in the definition of SNR because our maps are relatively small and can have non-zero means due to sample variance that vary from map to map, which can affect the consistency of the SNR definition. Note that the subtraction of $\mu$ does not affect $\sigma$, and it is not needed in the case of $\kappa$ reconstructed from the (directly observable) cosmic shear field. An example of a $\kappa$ map generated from numerical simulations through ray tracing is shown in Fig. \ref{fig:peaks and kappa map}.

For most of the paper, the lensing quantities are measured in maps obtained from cosmological simulations without added noise, except for Section \ref{section:GSN} where galaxy shape noise is included and all quantities are measured from the noise-added maps (for more details see Section \ref{section:GSN}).

\subsection{Numerical simulations} \label{section: Numerical Simulations}

In order to study weak lensing peak statistics, in this paper we use a large suite of WL convergence maps constructed from two sets of N-body simulations. The first are the publicly-available all-sky convergence maps of \citet{Takahashi2017} (hereafter \citetalias{Takahashi2017}). These maps have a source redshift of $z_s=1$ and have been generated using the ray tracing algorithm described in \citet{Haman2015} \citep[see also][]{Shirasaki2015}, with a HEALpix resolution of $N_{\rm{side}}=16384$. 
To avoid probing the same structures along the line-of-sight, \citetalias{Takahashi2017} constructed the light cone by stacking a hierarchy of cubic simulation boxes, with comoving sizes $L,2L,3L,\cdots,14L$, where $L=450h^{-1}$Mpc. The simulations had a particle number of $2048^3$, where the particle mass depends on the box size, and ranges from $8.2\times10^8$ to $2.3\times10^{12}M_\odot$ (see Table 1 of \citetalias{Takahashi2017} for more details). Each of the simulation boxes was duplicated 8 times and nested around the observer, such that nests of larger boxes contain nests of smaller boxes at their centers. Ray tracing was then performed on the nested simulation boxes by  
taking the projected mass distribution in spherical shells of $150 \hMpc$ in thickness centred on the observer (see \citetalias{Takahashi2017} for illustration). The cosmological parameters adopted for the \citetalias{Takahashi2017} simulations are $\Omega_m = 0.279$, $\sigma_8 = 0.820$ and $h = 0.7$, where $h = H_0 / 100$ km s$^{-1}$ Mpc$^{-1}$. Throughout this paper we have split the \citetalias{Takahashi2017} all-sky map into 184 separate \map{10} maps with $2048^2$ pixels per map, for which we can use the flat sky approximation to simplify our analysis. A detailed description of the method we use to split the all-sky map into smaller squares is given in Appendix \ref{Appendix: split all sky map}

The second set of WL maps we use are taken from \cite{Matilla2017} (hereafter \citetalias{Matilla2017}; see also \citealt{Gupta2018}) and consist of maps for 96 different cosmologies. It was built with the simulation pipeline described in \cite{Petri2016}.
For each cosmology, the maps were obtained from an N-body simulation of a periodic box with length $L = 240 \hMpc$ and $512^3$ simulation particles with a particle mass of $\sim10^{10}h^{-1}M_\odot$ (the exact value depends on the actual cosmology). Ray tracing was then performed by using a source redshift of $z_s = 1$ and by stacking particles into lens planes with a thickness of $80 \hMpc$ between the source and the observer. The lens planes were generated by taking a slice along a coordinate axis of the original simulation box and applying a random shift and rotation. This process was repeated to generate $512$ \map{3.5} maps per cosmology with $1024\times1024$ pixels per map. Note that each of the 512 maps were obtained from the same periodic simulation by varying the orientation of the line-of-sight direction. For a more detailed description we refer the reader to \citetalias{Matilla2017}. 

In total we have two sets of maps, one with $184$ \map{10} maps for a fixed cosmology and the other with $512$ \map{3.5} maps for 96 cosmologies with different values of $\Omega_m$ and $\sigma_8$. Larger maps are ideal for 2PCF studies as the 2PCF cannot be reliably calculated at large separations where pair measurements are affected by the finite size of the map. However, the differences in the two simulation data sets used here bring some benefits for our analysis. First, given the simulations use different ray tracing codes and box tiling methods, if we are able to identify certain features of the WL peak statistics in both simulations, this can be a check that the said features are not an unphysical consequence of the procedure used to generate the convergence maps. Second, the different simulation maps have different angular sizes and resolutions, which can help highlight any potential systemics in our analysis due to the box size or the pixel resolution.

\subsection{Weak lensing peaks}
WL peaks in this paper are defined as the maxima of the convergence field, which trace local over-densities in a range of environments. To extract the WL peaks, we first smooth the convergence map with a Gaussian filter. The convergence field has power on all scales, so the number and spatial distribution of WL peaks depends on the smoothing scale, with a larger smoothing washing out low contrast peaks and merging neighbouring peaks. We mainly study peaks identified with a Gaussian filter with smoothing length $\theta_s = 1\arcmin$, a range of smoothing scales have been studied in \cite{J.Liu2015}, showing that this smoothing scale is ideal for WL peak studies. In some cases (which will be explicitly mentioned), we vary $\theta_s$ to understand how the results depend on smoothing scale. Next we identify WL peaks by finding all pixels in the maps whose values are larger than those of their 8 neighbours, and peaks within $3 \theta_s$ of the map boundary are removed to avoid edge effects where the Gaussian filter is truncated.
The height of a peak is given by the $\nu$ value of the smoothed convergence field at the peak position. For a given convergence map, we can generate multiple peak catalogues by imposing a $\nucut$ threshold and keeping only peaks with $\nu \ge \nucut$.

Fig. \ref{fig:peaks and kappa map} illustrates the distribution of peaks (shown as green dots) for three different SNR thresholds,  $\nucut = 1, 2$ and $3$. To highlight the distribution of peaks on both small and large scales, we show peaks identified with a Gaussian smoothing scale, $\theta_s = 2\arcmin$; using a smaller $\theta_s$ value would result in many more peaks and make the graph less legible.

Fig. \ref{fig:peaks and kappa map} shows that peak catalogues with different $\nucut$ values trace different features of the convergence field. The catalogue with $\nucut = 1$ traces the over-dense regions of the convergence map, whilst avoiding the darker under-dense regions. In particular, many peaks seem to be arranged in a somewhat filamentary pattern. By increasing $\nucut$ to 2, we find that the resulting catalogue has a significantly lower number of peaks and the peaks are now more clustered. Most of these peaks are found in highly over-dense regions, with some small filamentary patterns still remaining. Finally, there are few peaks with $\nucut = 3$, but they show a high degree of spatial clustering and are located in the very over-dense regions of the map.

The description of Fig. \ref{fig:peaks and kappa map} above highlights two important features in the behaviour of WL peaks: the number of WL peaks and their clustering, which are respectively quantified by two commonly used statistics, the peak abundance and the peak two point correlation function (2PCF). The former is well studied and has been considered for many cosmological applications \citep[e.g.,][]{J.Liu2015,J.Liu2016,X.Liu2016,Shirasaki:2016twn,Shirasaki2017,Shan2018,Li2018,Wei2018}, whereas weak lensing 2PCFs are usually measured as shear-shear correlations \citep{Fu2008,Heymans2012,Kilbinger2013}, with very few studies directly focused on the peak 2PCFs \citep{Marian2013,Shan2014}. 

The two point correlation function measures the probability of finding two points (or in our case, WL peaks) at a given separation ($\theta$ for angular separations on the sky). It can also be interpreted as a measure of the excessive clustering of a distribution of points relative to the clustering of randomly distributed points. 
To estimate the 2PCFs, we use the Landy-Szalay estimator \citep{Landy1993}, which is a robust way of measuring 2PCFs, especially for small maps and low tracer number densities. Using this estimator requires a catalogue of randomly distributed points, whose role is to account for boundary effects and serves as a proxy for the volume (area in 2D) of the sample. The Landy-Szalay estimator is evaluated as  
\begin{equation}
    \xi_{\rm{LS}}(\theta) = 1 + \bigg(\frac{N_R}{N_D}\bigg)^2 \frac{DD}{RR} - \bigg(\frac{N_R}{N_D}\bigg) \frac{DR}{RR} \, , 
    \label{eq:LS estimator}
\end{equation}
where $N_D$ and $N_R$ are the numbers of data and random points and $DD$, $DR$ and $RR$ are the number of data-data, data-random and random-random pairs in bins $\theta \pm \delta \theta$. We calculate the 2PCFs by taking the average over many small maps (see description in section \ref{section: Numerical Simulations}). 
Since the maps are small, taking the average of the $\xi$ values measured for each map leads to biased results and we discuss this subtlety in detail in appendix \ref{Appendix: small map bias}. To obtain unbiased results, we calculate the average of the $DD$, $DR$ and $RR$ pair counts over all maps, and then we insert the average pair counts into Eq. \eqref{eq:LS estimator}.

\vspace{ -.4cm}
\section{Weak lensing statistics}\label{section: Weak lensing statistics}
As mentioned above, we are mainly interested in the one- and two-point statistics of WL peaks. In order to gain some first insight into the properties of these quantities, we use the large, $100\degSq$, maps from the \citetalias{Takahashi2017} simulations for the results shown in this section. In the next section we shall use the small maps from the \citetalias{Matilla2017} simulations to quantify the dependence of peak statistics on cosmology.

\subsection{Peak abundance} \label{section: T17 PA} 

\begin{figure}
    \centering
    \includegraphics[width=\columnwidth]{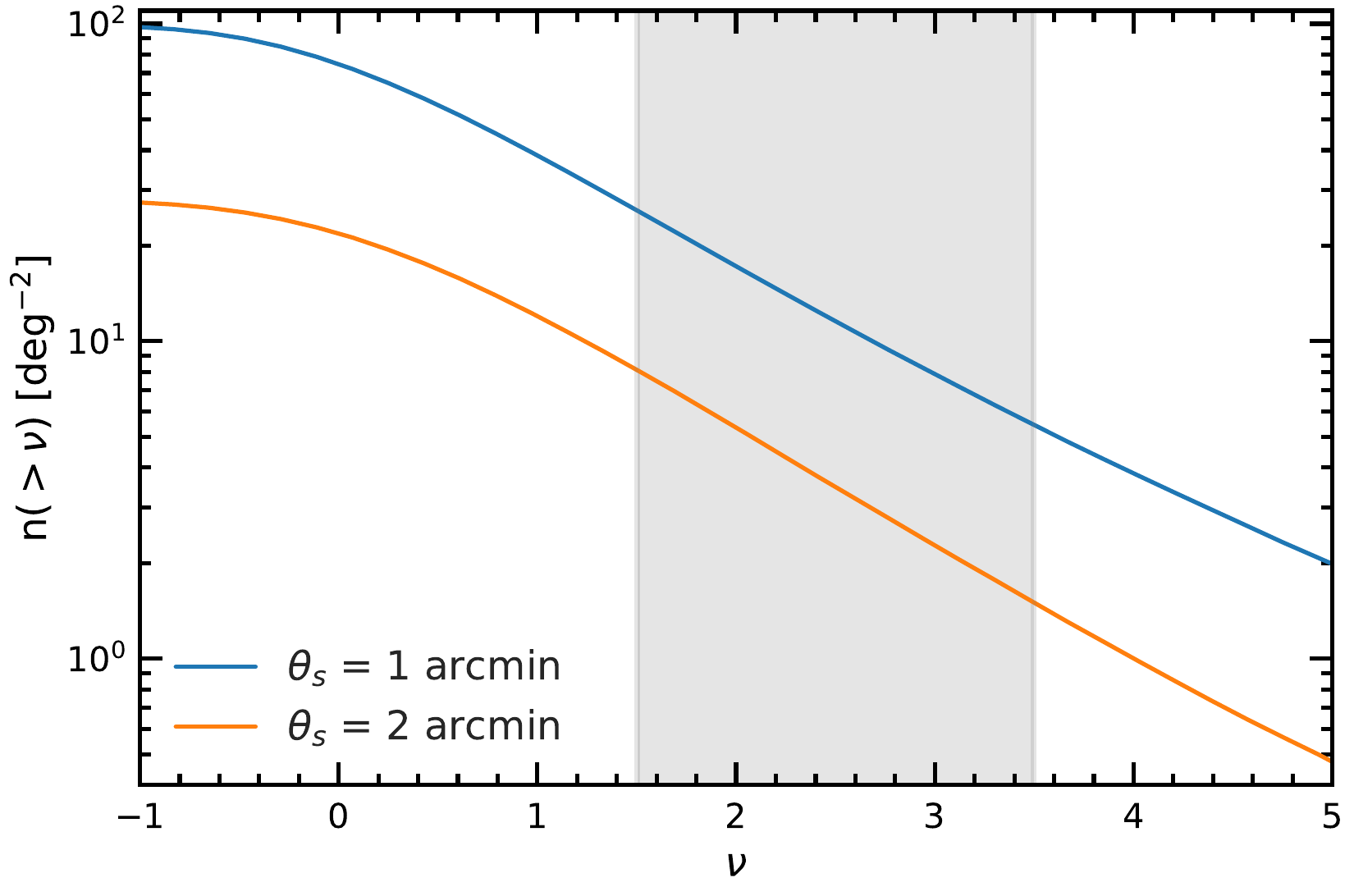}
    \vskip -.2cm
    \caption{WL peak number density as a function of peak signal-to-noise, $\nu$, for two smoothing scales, $\theta_s = 1\arcmin$ (blue) and $\theta_s = 2\arcmin$ (orange) from the \citetalias{Takahashi2017} maps. The shaded grey region highlights the $\nu$ range that we study in this paper, discussed in section \ref{section: 2PCF rescaling and self-similarity}. 
    }
    \label{fig:peaks abundance}
\end{figure}

We start by studying the mean abundance of WL peaks, which is expressed in terms of the cumulative peak abundance, $n(>\nu)$. This represents the number density in deg$^{-2}$ of all peaks whose SNRs are higher than $\nu$. The peak abundance is illustrated in Fig.~\ref{fig:peaks abundance}, where the results are averaged over the 184 \citetalias{Takahashi2017} maps. The blue (upper) and orange (lower) curves correspond to a Gaussian smoothing kernel $\theta_s=1$ and $2\arcmin$, respectively. According to Eq.~(\ref{Eq:nu}), the smoothing scale $\theta_s$ enters the definition of $\nu$ in two ways, by affecting the pixel values of $\kappa$ and the overall rms $\kappa$ fluctuation, $\sigma$. For the \citetalias{Takahashi2017} maps we find $\sigma=0.013$ and $0.010$ respectively for $\theta_s=1$ and $2\arcmin$, and in the next section we will see that $\sigma$ has a clear cosmology dependence as well.

The qualitative behaviour shown in Fig.~\ref{fig:peaks abundance} is as expected. There are very few peaks with high $\nu$ values since these correspond to massive dark matter structures, which are rare. As $\nu$ decreases, the peak abundance, $n(>\nu)$, increases quickly until $\nu\sim0$ since lower $\nu$ values correspond to lower mass and thus more abundant dark matter structures. However, for $\nu\lesssim0$\footnote{Note that WL peaks can have $\nu<0$, or equivalently $\kappa<0$. These are local maxima in underdense regions of the convergence map.} we see that $n(>\nu)$ flattens, showing that there are few peaks with $\nu<0$. It highlights that there are few structures in underdense regions that are massive enough to lead to a local maximum, especially when smoothing over $1$ and $2\arcmin$. Increasing the smoothing scale $\theta_s$ leads to a lower peak abundances at fixed $\nu$, since smoothing over a larger region tends to eliminate some peaks. 

\begin{figure*}
    \centering
    \includegraphics[width=2\columnwidth]{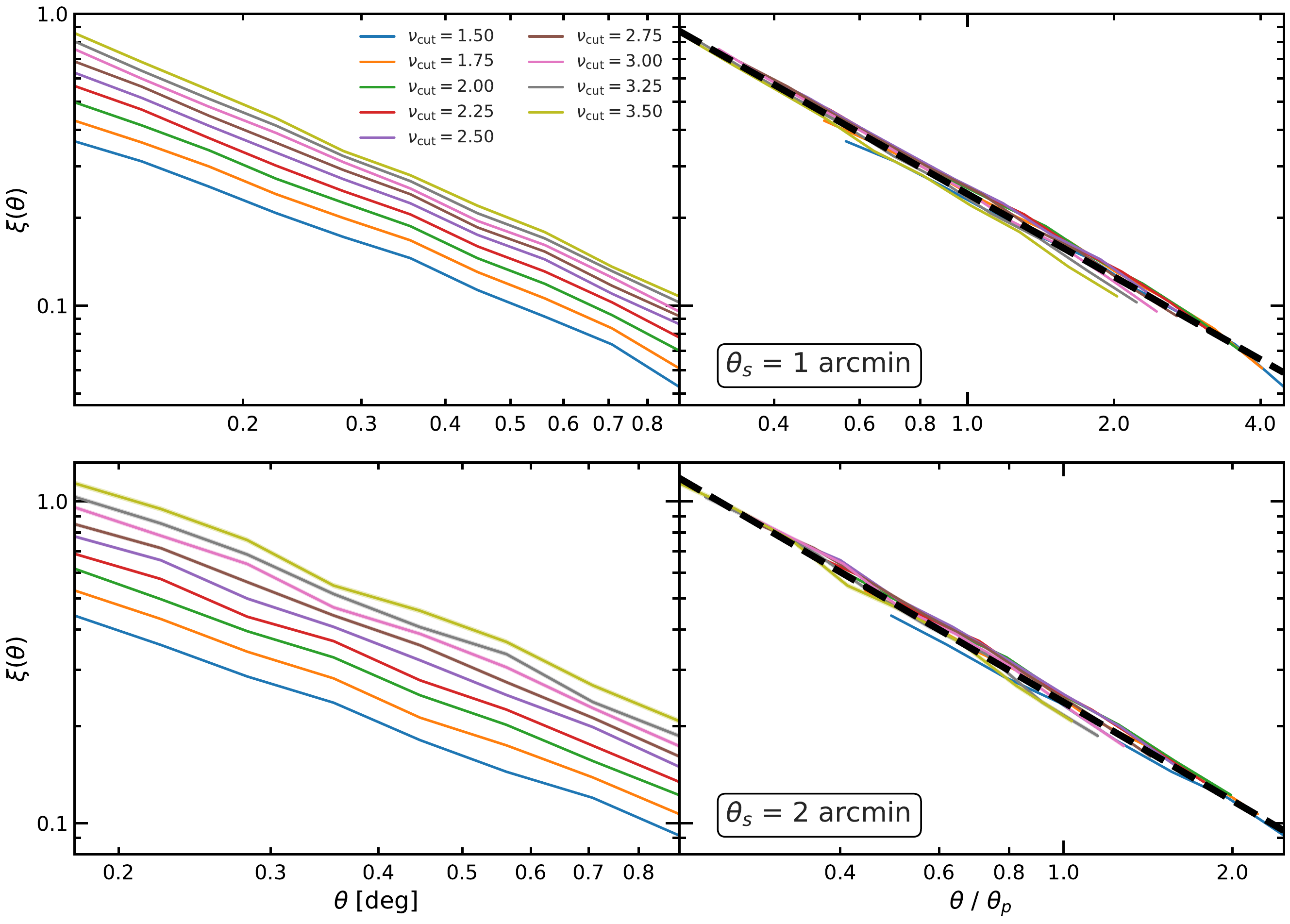}
    \vskip -.3cm
    \caption{The 2PCFs of WL peaks for different peak catalogues obtained by varying the threshold, \nucut{}, in the range $[1.5,3.5]$ in increments of $\Delta \nucut = 0.25$. The results are for the \citetalias{Takahashi2017} maps and for two different smoothing scales, $\theta_s = 1$ (top-row) and $2\arcmin$ (bottom-row). The left column shows the 2PCFs as a function of angular separation, $\theta$. The right hand column shows the rescaled 2PCFs,which are the 2PCFs expressed in terms of $\theta/\theta_p$, with $\theta_p$ the mean peak separation in the catalogue. The 2PCFs displays a striking self-similar behaviour, with all rescaled 2PCFs curves lying on top of each other. The black dashed line in the right-hand column shows the best fitting power law to the rescaled 2PCFswith gradients $-0.94$ (top) and $-1.02$ (bottom). 
    }
    \label{fig:self sim 1and2}
\end{figure*}

We have checked that the WL peak abundance shown in Fig.~\ref{fig:peaks abundance} can be well fitted by the function $n({>}\nu) = -a \left[\tanh(b\nu)-1\right]$ for the entire $\nu$ range. However, for reasons that will become clear in Section \ref{section: 2PCF rescaling and self-similarity}, in this paper we are interested in the range $\nu\in[1.5,3.5]$ (shown as the grey shaded region in Fig.~\ref{fig:peaks abundance}), where $n(>\nu)$ can be modelled  as a power law (see Section \ref{section: Peak abundance model}). 
Note that Fig.~\ref{fig:peaks abundance} also shows the uncertainties in the $n(>\nu)$ measurement, which are the standard errors of the mean of the 184 \citetalias{Takahashi2017} maps; however, these error bars are not visible as they are roughly of the same size as the line width.

\subsection{Peak two point correlation function and $\nu_{\rm{cut}}$ dependence} 
\label{section: T17 2PCF}

For WL peaks it has been suggested that the 2PCF can be well modelled by a power law \citep{Shan2014}. In this section, we will confirm this power-law description using the \citetalias{Takahashi2017} convergence maps, and show that it works well for peak catalogues with a wide range of $\nucut$ thresholds.

The 2PCF dependence on $\nucut$ is of particular interest, because by decreasing $\nucut$ we are including lower peaks into the analysis, which is equivalent to incorporating smaller dark matter structures into the clustering statistics. In the current standard cosmological paradigm, large-scale structures (LSS) evolve hierarchically, with larger objects forming from higher initial density peaks. This means that by varying $\nucut$ we probe the different regimes of nonlinear LSS formation and thus potentially provides more powerful cosmological tests. As an example, in certain modified gravity models, smaller structures experience a stronger boost in their nonlinear growth \citep[e.g.,][and references therein]{Clifton2012}, and we expect this to leave potentially detectable signatures in the peak 2PCFs at different $\nu_{\rm cut}$ values. In addition, as we have seen above, lowering the $\nucut$ threshold increases the number of peaks included in the catalogue, and this can help increase the statistical constraining power. We will see shortly that there is a self-similarity in the peak 2PCF, which means that having peak catalogues for multiple \nucut{} values does not require separate modelling for each catalogue; this can potentially strongly improve the constraining power by WL peaks. 

The left panels of Fig.~\ref{fig:self sim 1and2} show the mean 2PCFs of the \citetalias{Takahashi2017} maps for a range of \nucut{} values and for two smoothing scales, $\theta_s=1$ and $2\arcmin$. The error bars, which are the standard errors of the 184 maps, are shown as shaded regions around the curves, but they are very small and barely visible.

A quick inspection of Fig.~\ref{fig:self sim 1and2} by eye confirms that the 2PCFs are well described by power laws. We can see that as $\nu_{\rm{cut}}$ increases the amplitude of the 2PCF, $\xi(\theta)$, also increases. This is intuitive to understand: the high WL peaks correspond to more massive structures which tend to cluster more strongly. Moreover, the 2PCF amplitude is higher for peak catalogues obtained using a larger smoothing length, $\theta_s$. This is because a higher $\theta_s$ value suppresses peaks originating from low mass dark matter structures that cluster less. The gradient of the 2PCFs from maps with a fixed $\theta_s$ increases slightly with $\nucut$, but this effect is weak for both smoothing scales shown, and the dominant effect of varying $\nucut$ is in the amplitude of the 2PCF.

Note that smoothing can lead to a merging of peaks separated by distances comparable to the smoothing scale $\theta_s$, and therefore eliminates some peaks which are close to each other. This leads to a drop off of the 2PCF from the power law on scales $\lesssim\theta_s$, which is why we show a different $\theta$ range in the two left panels of Fig.~\ref{fig:self sim 1and2}. Additionally, for a given WL map size, the 2PCFs cannot be reliably measured at large separations as there are too few peak pairs, which is why in Fig.~\ref{fig:self sim 1and2} we adopted a conservative $\theta_{\rm max}$ which is 1/10 the map size. Therefore, the smoothing scale and the map size set a limited range in $\theta$ within which we can measure the 2PCF. More explicitly, while for the \citetalias{Takahashi2017} maps we use $\theta_s\in[1,3]$ arcmin in this study, for the smaller \citetalias{Matilla2017} maps we only use $\theta_s=1$ arcmin to avoid having a too narrow $\theta$ range. A larger $\theta_s$ is necessary for maps where galaxy shape noises (GSNs) are included, to suppress the biasing effects caused by the latter \citep[e.g.,][]{Davies2018}; we will come back to this point in Section \ref{section:GSN} below.

\subsection{2PCF rescaling and self-similarity}
\label{section: 2PCF rescaling and self-similarity}

We now move on to one of the most important results of this work: the self similarity of the peak 2PCFs. This has been first studied (very briefly) in \citet{Davies2018} in the context of explaining the self-similar behaviour of the abundances for voids identified from WL peaks with varying $\nucut$. As we show later, the 2PCF self similarity is a very useful property that merits the more detailed investigation presented here.

The quest for a self-similar behaviour in the peak 2PCFs is motivated by the following observations: the 2PCF amplitude is lower for peak catalogues with lower $\nucut$; meanwhile, these catalogues have more peaks and hence a smaller mean peak separation, $\theta_p$. By expressing the 2PCF in terms of $\theta/\theta_p$ 
the various curves could potentially be brought closer together. The question is whether after this rescaling the 2PCF curves for different $\nucut$ thresholds can be made to overlap, in which case their modelling can be significantly simplified.

The right panels in Fig. \ref{fig:self sim 1and2} show the rescaled 2PCF, that is the 2PCF expressed as a function of $\theta/\theta_p$ instead of $\theta$. To obtain this result, we calculated the mean peak separation as $\theta_p = (N/A)^{-1/2}$, where $N$ is the number of peaks in a catalogue and $A$ is the area of the map. The $\theta_p$ value for a peak catalogue can be directly inferred from the peak abundance, $n(>\nu)$, as $\theta_p = n(\nu > \nucut)^{-1/2}$. We find that the rescaled 2PCFs lie on top of each other and thus it indicates that the peak 2PCF is self similar. This shows that the one-point statistic of WL peaks, $n(>\nu)$, can be tied to the amplitude of the 2PCF to achieve the mentioned self similarity. The self-similar behaviour is mainly limited to the range $\nucut \in [1.5,3.5]$, with the rescaled 2PCFs starting to peel off from the average relation for $\nu_{\rm{cut}} < 1.5$ and $\nu_{\rm{cut}} > 3.5$. At this stage it is unclear whether the breakdown of self-similarity at $\nucut>3.5$ is physical or due to the small number of high SNR peaks in our maps (which, as discussed in Appendix \ref{Appendix: small map bias}, could bias the estimation of the two-point correlation function); this will be investigated in more detail in the future. Bearing this issue in mind, in this work we limit our investigation to the modelling of WL peak statistics for $1.5 < \nucut < 3.5$ only.
Note that this happens to be the same range within which the peak abundance can be well described by a power law (see Section \ref{section: T17 PA}).

The self-similar behaviour holds for both smoothing scales shown in Fig.~\ref{fig:self sim 1and2}, however, the rescaled 2PCFs for the larger smoothing length ($\theta_s=2\arcmin$; bottom right panel of Fig.~\ref{fig:self sim 1and2}) are shifted to lower $\theta / \theta_p$ values than the results for $\theta_s=1\arcmin$. It suggests that 2PCFs are self-similar when keeping the smoothing scale constant, and that the self-similarity behaviour does not hold when comparing 2PCFs obtained for peak catalogues with different smoothing scales. 

The panels in the right column of Fig.~\ref{fig:self sim 1and2} also show that the rescaled 2PCFs are well fitted by a power law, as shown by the black dashed curves with gradients $-0.94$ ($\theta_s = 1\arcmin$) and $-1.02$ ($\theta_s = 2 \arcmin$) .

\begin{figure}
    \centering
    \includegraphics[width=\columnwidth]{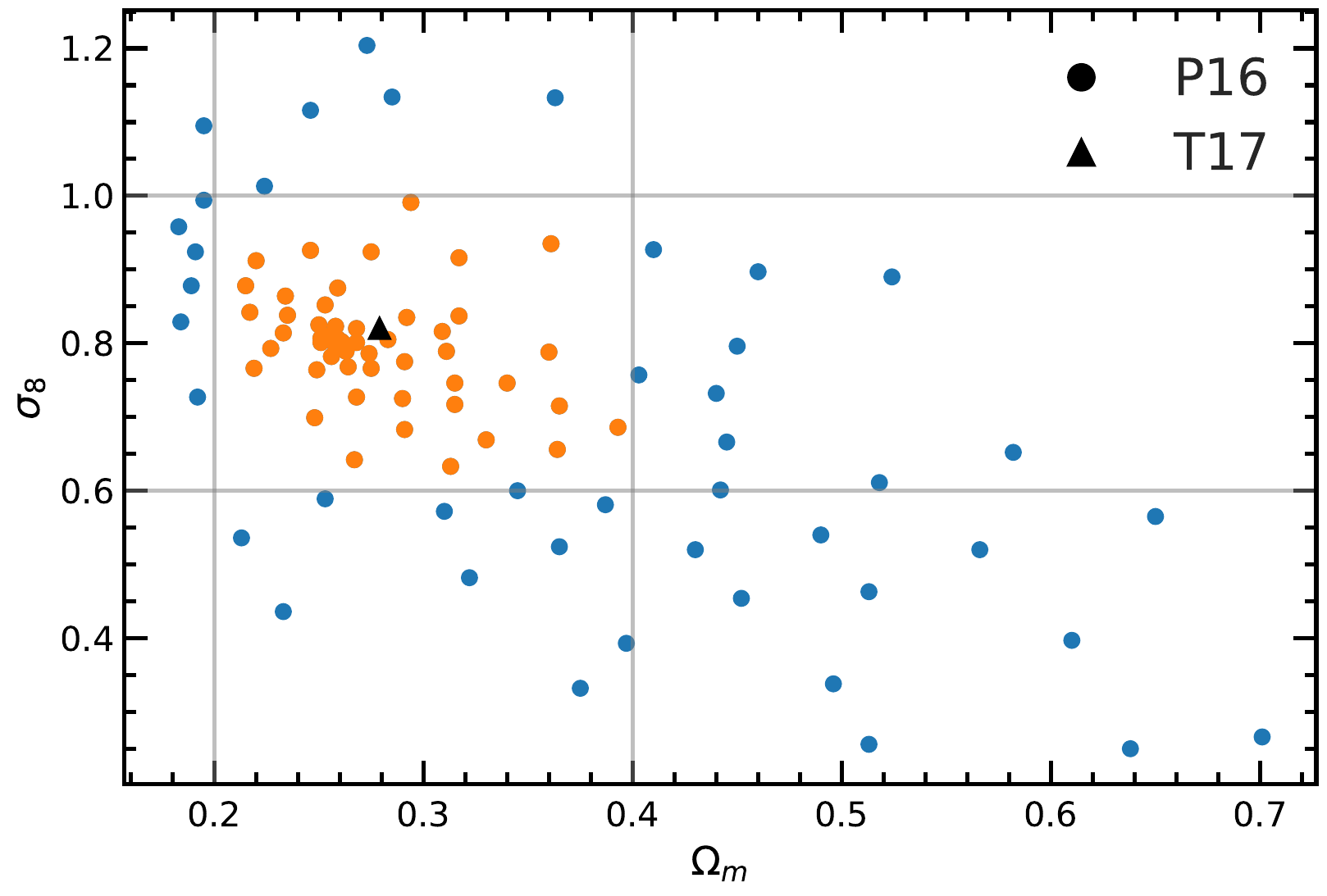}
    \vskip -.3cm
    \caption{The $(\Omega_m$,$\sigma_8)$ parameter space that is probed with our suite of 96 simulations from \citetalias{Matilla2017}. The faded grey lines indicate the cuts that are made to remove extreme cosmological parameters and give the orange points which we use to construct our $\Omega_m$,$\sigma_8$ dependant model for 2PCF reconstruction in section \ref{section: cosmo dependence}. The black triangle shows the $\Omega_m$ and $\sigma_8$ values of the \citetalias{Takahashi2017} simulations.
    }
    \label{fig:cosmo models}
\end{figure}

\begin{figure*}
        \includegraphics[width=\columnwidth]{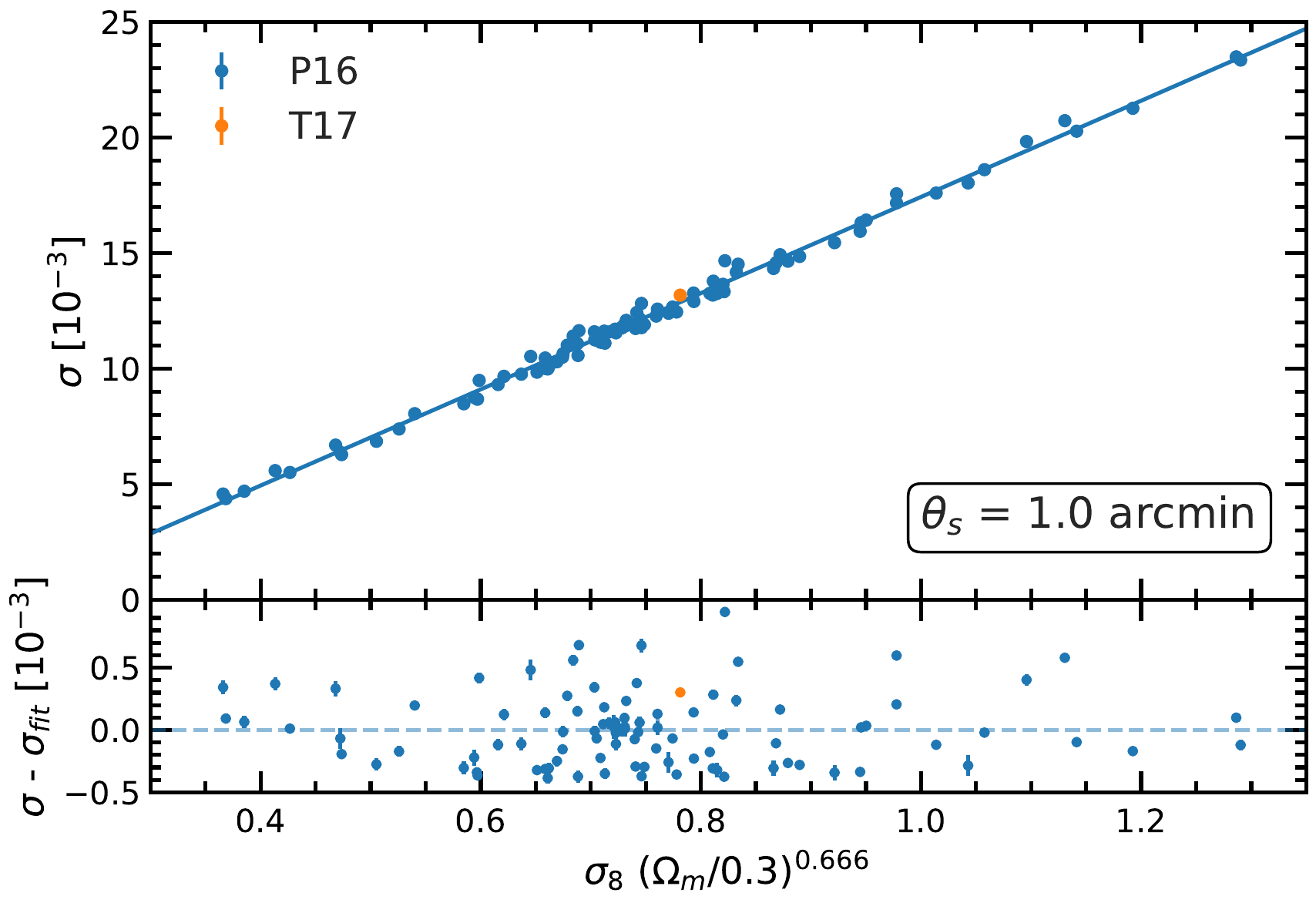}
        \includegraphics[width=\columnwidth]{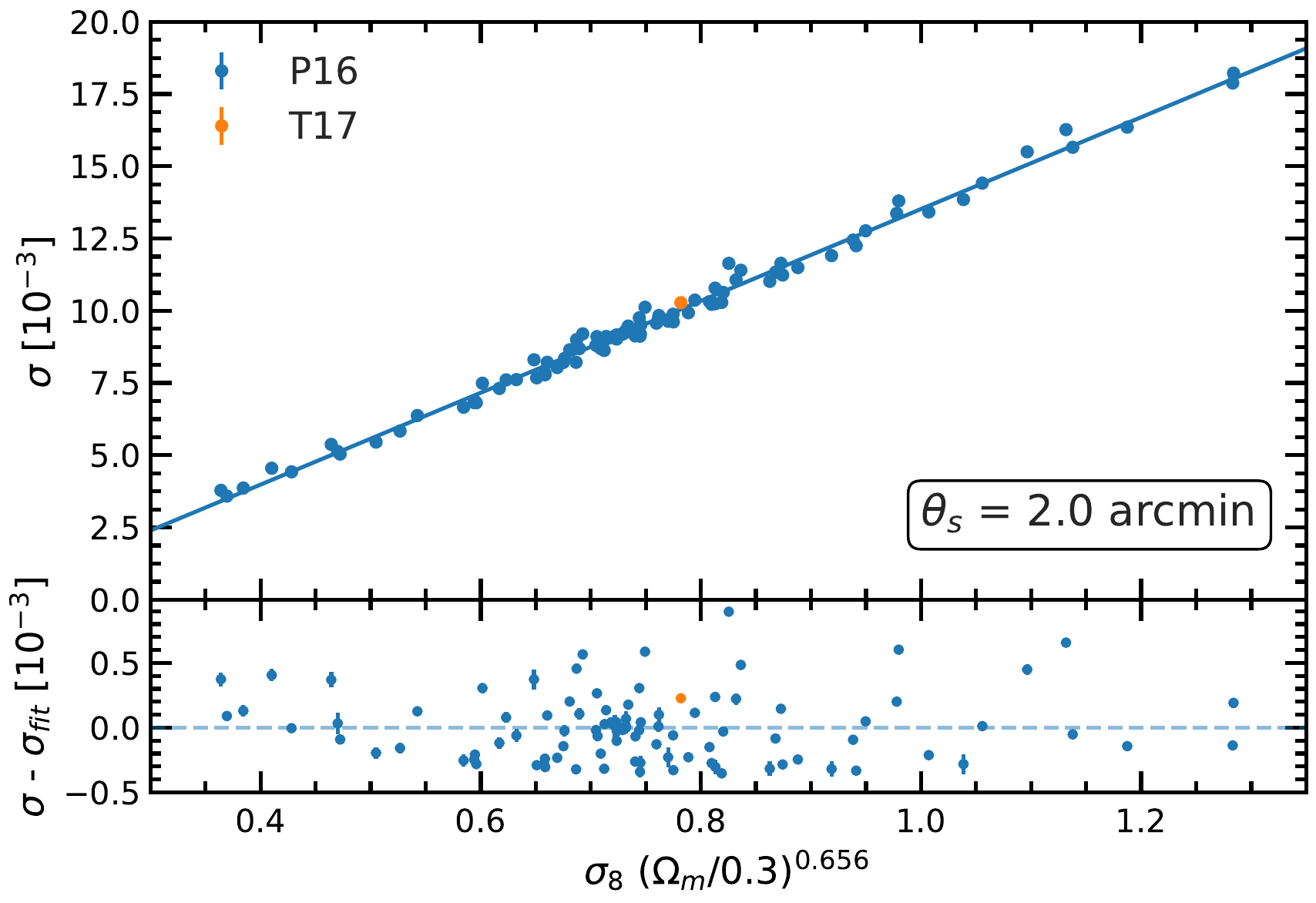}
    \caption{The rms fluctuations, $\sigma$, of the WL convergence map as a function of $\sigma_8(\Omega_m/0.3)^{\alpha}$. The power $\alpha$ indicates the degeneracy direction between $\sigma_8$ and $\Omega_m$ that gives the same rms fluctuations in the convergence field. The blue points correspond to the 96 cosmologies from \citetalias{Matilla2017} (see Fig. \ref{fig:cosmo models}) and the orange point corresponds to the \citetalias{Takahashi2017} one. The left and right hand panels show the mean $\sigma$ values for two smoothing scales, $\theta_s$ = 1 and 2 arcmin, respectively. The bottom sub-panels show the residuals between the mean $\sigma$ values and the best fitting line (blue). The vertical bars show the standard errors, which may be underestimated as discussed in Appendix \ref{Appendix: error estimation}.}
    \label{fig:std cosmo}
\end{figure*}

\section{Cosmology dependence and universal fitting functions}\label{section: cosmo dependence}

In this section we study the dependence of peak abundance and peak 2PCF on the $\Omega_m$ and $\sigma_8$ cosmological parameters by analysing these statistics for the set of 96 different cosmologies used for the \citetalias{Matilla2017} maps. The ($\Omega_m$,$\sigma_8$) parameter space of the \citetalias{Matilla2017} maps is indicated by the points in Fig. \ref{fig:cosmo models}. The parameter space is densely sampled around $\Omega_m=0.26$ and $\sigma_8=0.8$, which corresponds to the fiducial cosmology, and only sparsely sampled for models with very different parameter values. In particular, when describing the cosmology dependence of various peak statistics, we will limit our fitting procedure to the ($\Omega_m$, $\sigma_8$) pairs shown as orange points in the figure. This removes extreme and unrealistic cosmological parameters from our analysis. For comparison, the parameters used for the \citetalias{Takahashi2017} maps are indicated by the black triangle in Fig.~\ref{fig:cosmo models}.

The two cosmological parameters, $\Omega_m$ and $\sigma_8$, are degenerate because they can impact the size of the matter fluctuations in similar ways, and the direction of degeneracy depends on the physical quantity which is being studied. In order to better assess the potential and limitations of using WL peak statistics to constrain these cosmological parameters, it is important to know the degeneracy direction for the physical quantities of interest. Following the usual approach, we define the parameter combination, $\Sigma_8(\alpha)\equiv\sigma_8\left(\Omega_m/0.3\right)^\alpha$, where $\alpha$ characterises the degeneracy for a given statistic ($\alpha$ is allowed to vary for different statistics since they usually do not have exactly identical degeneracy directions). Note that for studying the cosmology dependence we use only the \citetalias{Matilla2017}, and not the \citetalias{Takahashi2017}, convergence maps, and the latter is used as a consistency check of the fitted models.

The fittings carried out in this section are mainly to exemplify the cosmology dependence of the self-similar feature present in the 2PCF in Fig.~\ref{fig:self sim 1and2}, which applies only to theoretical (simulated) lensing maps with no noises and with a specific smoothing length. 
Before this approach can be used for observational constraints, further development will be required, notably the inclusion of galaxy shape noise. We discuss briefly the impact the latter has on the self similarity of the 2PCF in Section \ref{section:GSN}. In order to study the rescaled 2PCF for a range of cosmologies using more realistic noise-added maps, larger simulations for this range of models are required, which we leave to future work.

\subsection{Convergence rms fluctuation}
\label{section: Convergence map standard deviation}

We describe peaks in terms of the convergence SNR value at their position. To calculate this, we use the root-mean-square (rms) fluctuations, $\sigma$, of the convergence field (see e.g. Eq.~\ref{Eq:nu}). In principle, $\sigma$ is used merely as a normalisation factor and it is not entirely unreasonable to use the same value to define $\nu$ across all cosmologies. However, the standard deviation (or rms fluctuation) of the corresponding WL convergence map, $\sigma$ is a quantity with a clear physical meaning, and hence it is natural to use its correct value for a given cosmology. Therefore, we need a general description of $\sigma$ as a function of input cosmological parameters, $\sigma=\sigma(\Omega_m,\sigma_8)$. Having this function is also of interest on its own, since it is useful to know how the rms fluctuation of the WL convergence field depends on the cosmological model.

The dependence of the convergence rms fluctuation, $\sigma$, on $\sigma_8$ and $\Omega_m$ is illustrated in Fig.~\ref{fig:std cosmo}, where we show the results for two smoothing lengths, $\theta_s=1$ (left panel) and $2\arcmin$  (right panel). In both cases we varied $\alpha$ such that $\sigma$ is well described by a linear function of $\Sigma_8(\alpha)$, that is:
\begin{equation}
    \sigma = m \Sigma_8(\alpha)  + c
     \equiv m \sigma_8 \bigg( \frac{\Omega_m}{0.3} \bigg) ^{\alpha}  + c
    \label{Eq: sigma relation} \;.
\end{equation}
We achieved this by performing a $\chi^2$ minimisation procedure with three free parameters: $m$, $c$ and $\alpha$. The best fitting parameter values are:
$m = 2.08\times10^{-2}$, $c = -3.39\times10^{-3}$ and $\alpha = 6.66\times10^{-1}$ for $\theta_s = 1\arcmin$ and $m = 1.59\times10^{-2}$, $c = -2.37\times10^{-3}$ and $\alpha = 6.56\times10^{-1}$ for $\theta_s = 2\arcmin$. For these fits, we used the standard errors in the determination of $\sigma$. These errors are shown in Fig.~\ref{fig:std cosmo} as vertical error bars, however in most cases the error bars are smaller than the size of the data points plotted and are not visible. The bottom panels in Fig.~\ref{fig:std cosmo} show the residuals, i.e., the deviations of the $\sigma$ values for each cosmology from the best-fitting lines. The fit residuals are small when compared to the absolute values of $\sigma$ and show no systematic trend with $\Sigma_8$, indicating that Eq.~\eqref{Eq: sigma relation} provides an excellent description for both smoothing lengths. In particular, we notice that the best-fit value of $\alpha$ is similar for $\theta_s=1$ and $2\arcmin$, and thus weakly dependent on the smoothing scale,. It is also reassuring to note that \citetalias{Takahashi2017} is in very good agreement with the best-fitting lines, which are obtained from the \citetalias{Matilla2017} convergence maps only, even though the two sets of maps are constructed from very different simulations.

Even though it is not the main line pursued in this work, we note that the measured standard deviation of the (reconstructed) WL maps is a useful quantity \citep[see, e.g.,][]{VanWaerbeke2013} and therefore a simple fitting formula for $\sigma$ in terms of $\Omega_m$ and $\sigma_8$ will be useful both theoretically and observationally. 
However, because GSN introduces a major systematic uncertainty in real WL maps, it is necessary to study the  Eq.~\eqref{Eq: sigma relation} fitting formula using maps in which realistic GSN is included; this is beyond the scope of this study, because the \citetalias{Matilla2017} maps used in the analysis above have relatively small sizes. In this paper, instead, the primary use of Eq.~\eqref{Eq: sigma relation} will be to define $\nu$ for a convergence map with given $\Omega_m$ and $\sigma_8$ values.

As shown in Figure 5, within the large range of cosmological parameters covered in this study, $\sigma$ varies strongly, by up to a factor of $5$-$7$. By defining $\nu$ relative to the $\sigma$ of the corresponding model, we are able to define the SNR in different models relative to their own clustering amplitude. The alternative way to is use a constant $\sigma$ definition, such as the value for a fiducial model or the rms of the typical noise map. However, in our case this would mean comparing the clustering of map pixels with smaller $\kappa$ values in one model to the clustering of pixels with large $\kappa$ values in another model, when using the same $\nu_{\rm{cut}}$, and so a cosmology dependant $\nu_{\rm{cut}}$ range would have to be applied which is a far more complicated approach. We have explicitly checked by defining the SNR $\nu$ in Eq.~(\ref{Eq:nu}) using the $\sigma$ of the fiducial model where $\sigma_8=0.8$ and $\Omega_m=0.3$, and found that the self-similar behaviour still holds though slightly worse than shown here. Later in Section \ref{section:GSN}, when dealing with noisy maps, we shall use the $\sigma$ measured from the smoothed noisy maps to define the SNR.

\subsection{WL peak abundance}
\label{section: Peak abundance model}

\begin{figure}
    \centering
    \includegraphics[width=\columnwidth]{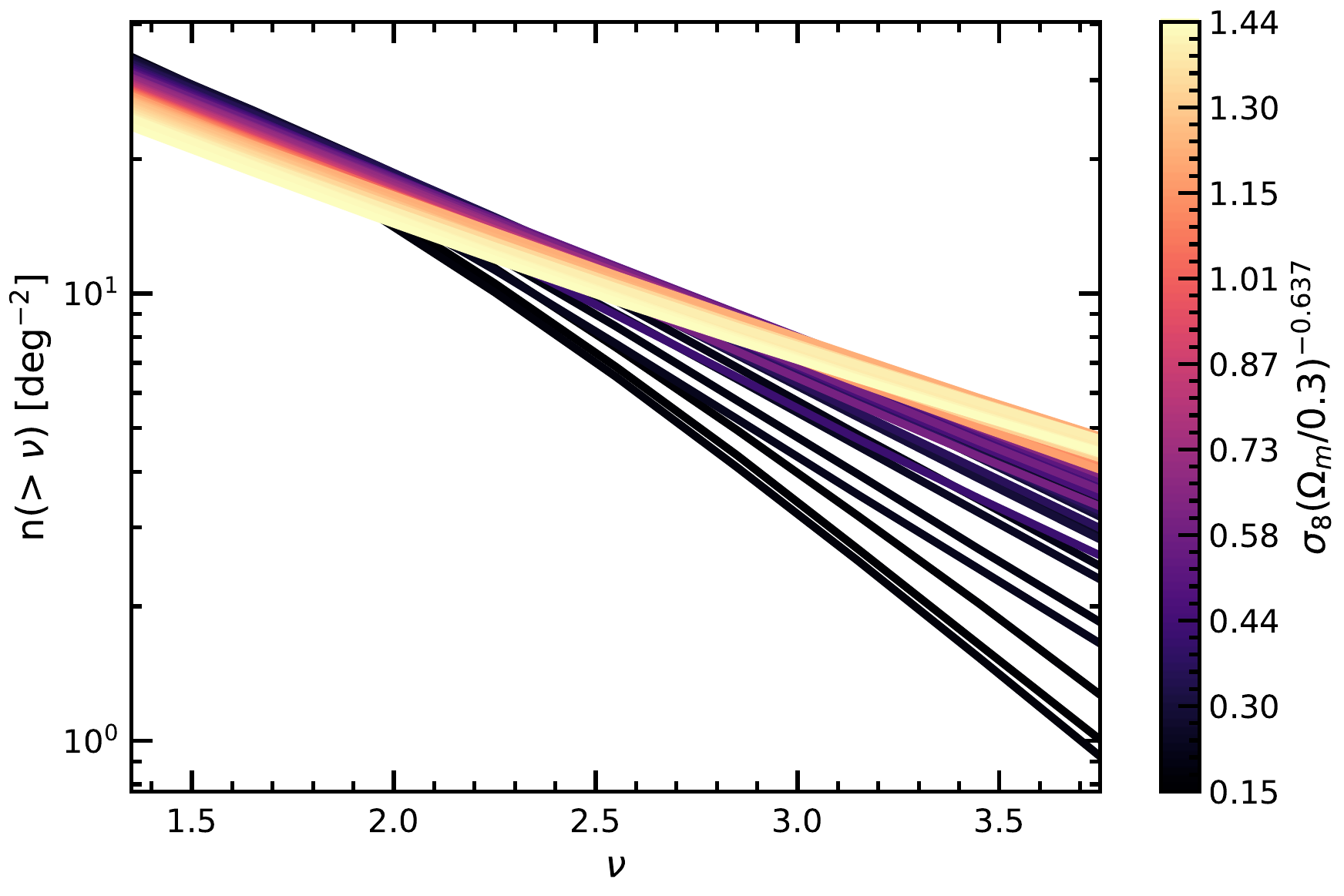}
    \vskip -.3cm
    \caption{The peak abundance, $n(>\nu)$, for the 96 cosmologies from \citetalias{Matilla2017} plotted against $\nu$. The lines are coloured according to a combination of the cosmological parameters $\Omega_m$ and $\sigma_8$ (see colour bar). 
    }
    \label{fig:all PA}
\end{figure}

\begin{figure*}
    \centering
    \includegraphics[width=\textwidth]{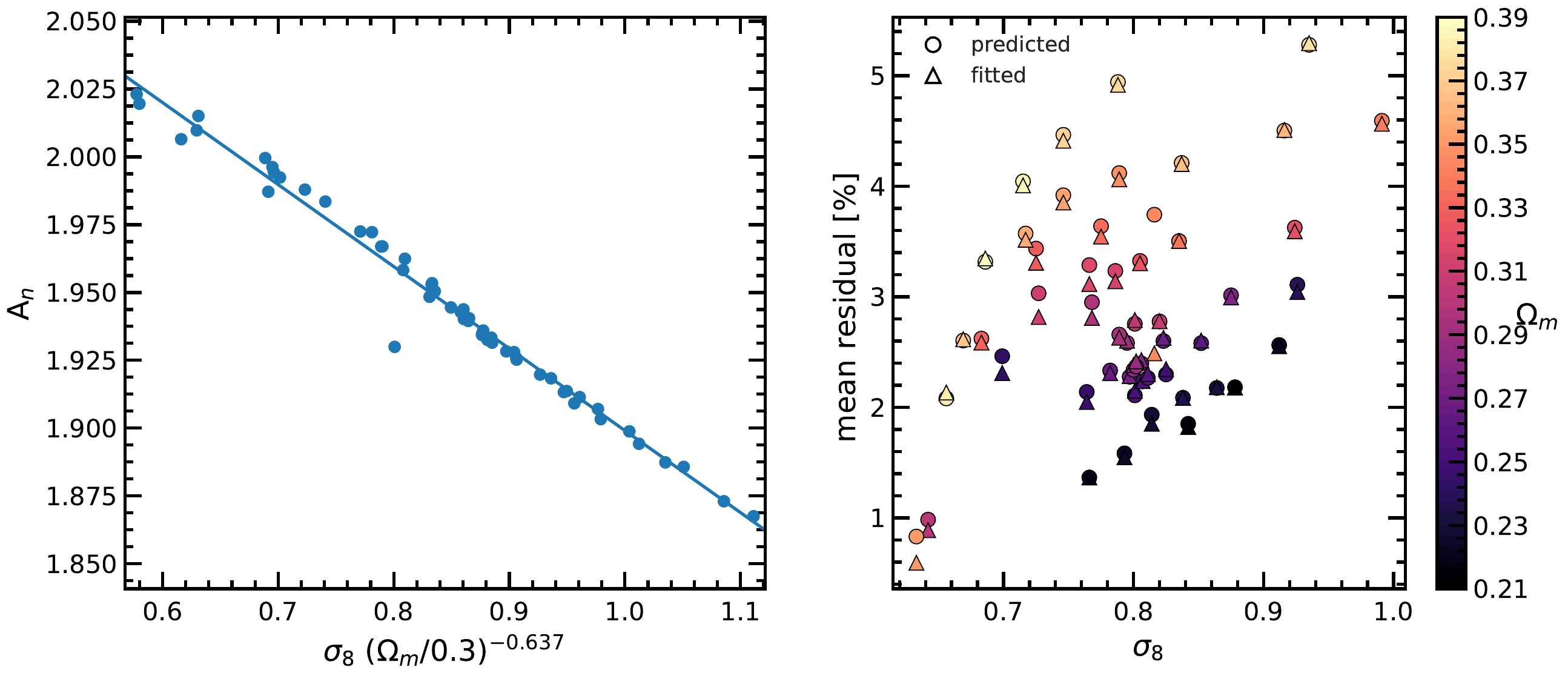}
    \vskip -.3cm
    \caption{ 
    \textit{Left panel:} 
    the dependence of $A_n$, which is the one parameter fit to the peak abundance (see Eq. \ref{eq:PA power law}), on a combination of $\sigma_8$ and $\Omega_m$ parameters. The exact combination is $\sigma_8(\Omega_m / 0.3)^{-0.637}$, where the power represents the degeneracy direction between $\sigma_8$ and $\Omega_m$ that gives the same $A_n$ value. The blue line shows the best fitting linear function (see Eq. \ref{Eq: a - one param PA}).
    \textit{Right panel:} 
    the triangle symbols show the percentage residuals of the one parameter power law fit to the peak abundance. The circle symbols show the extent to which our model can predict the peak abundance. The model works in two steps: i) use the blue solid line shown in the left panel to predict $A_n$ for a pair of $\sigma_8$ and $\Omega_m$ values, and ii) given $A_n$, infer the peak abundance using Eq. \eqref{eq:PA power law}. The various symbols are coloured according to the $\Omega_m$ cosmological parameter (see colour bar on the right).
    }
    \label{fig:PA cosmo and stats}
\end{figure*}

The peak abundances for the 96 models in the \citetalias{Matilla2017} simulations are shown in Fig.~\ref{fig:all PA}. The colour-bar shows the cosmological parameters for a given curve with the form $\Sigma_8 = \sigma_8(\Omega_m/0.3)^{-0.637}$. The spread of the amplitudes of the peak abundances across the 96 cosmologies is up to a factor of two. Most of the curves appear straight, indicating that the peak abundance is well described by a power law. The only exceptions are the black curves with very small values of $\Sigma_8$: these curves will be removed in the analysis as they correspond to the extreme cosmological parameter values indicated by the blue points in Fig.~\ref{fig:cosmo models}. As $\Sigma_8$ increases, the slope of the peak abundance decreases, while its amplitude increases at larger $\nu$ ($\nu\gtrsim2.5$) and decreases for small $\nu$ ($\nu\lesssim2.5$). This `rotation' of the curves about $\nu\approx2.5$ as $\Sigma_8$ changes implies a correlation between the slope and amplitude of the peak abundances as the cosmological parameters vary, and we shall see shortly that this fact can be utilised to reduce the number of fitting parameters in our peak abundance model.

In the range $1.5 < \nu_{\rm{cut}} < 3.5$, the peak abundance as a function of $\nu$ is well described by the power-law,
\begin{equation}
    \log n(>\nu) = B_n\nu + A_n,
    \label{eq:PA power law1}
\end{equation}
in which the fitting parameters $A_n$ and $B_n$ depend on the input cosmology.

In order to model $A_n = A_n(\Omega_m,\sigma_8)$ and $B_n= B_n(\Omega_m,\sigma_8)$, we fit Eq.~\eqref{eq:PA power law1} to the peak abundance for each of the cosmological models indicated by the orange points in Fig.~\ref{fig:cosmo models}. The fitting results confirm that $A_n$ and $B_n$ -- which respectively characterise the amplitude and slope of the peak abundance -- are strongly correlated, so that $B_n$ can be replaced with a function of $A_n$, $B_n(A_n)$, and the peak abundance can now be described using a one-parameter power law of the form
\begin{equation}
    \log n(>\nu) = \nu B_{n}(A_{n}) + A_{n},
    \label{eq:PA power law}
\end{equation}
with
\begin{equation}
    B_{n}(A_n) = -0.33A_n + 0.28.
    \label{eq:PA power law b(a)}
\end{equation}

This indicates that a universal model for WL peak abundance that works for a wide range of cosmological models can be obtained if one can fit the cosmology dependence of the single parameter $A_n$. The result is shown in the left panel of Fig.~\ref{fig:PA cosmo and stats}, where the $A_n$ values measured from the 96 \citetalias{Matilla2017} cosmologies are plotted against $\Sigma_8(\alpha)$ with $\alpha=-0.637$. The value of $\alpha$ corresponds to the one for which $A_n$ is well fitted by a linear function of $\Sigma_8(\alpha)$. The latter is shown as the solid line in the left panel of Fig.~\ref{fig:PA cosmo and stats}, and is given by 
\begin{equation}
    A_n = -0.30 \; \sigma_8(\Omega_m / 0.3)^{-0.637} + 2.20 \, .
    \label{Eq: a - one param PA}
\end{equation} 
For fitting the above equation the errors in $A_n$ are given by the uncertainties of fitting Eq.~\eqref{eq:PA power law b(a)} to the peak abundance. These errors are small and are not visible in Fig.~\ref{fig:PA cosmo and stats} since they are smaller than the symbol size.

Eqs.~\eqref{eq:PA power law}-\eqref{Eq: a - one param PA} can be used to predict the peak abundance for any input $\Omega_m$ and $\sigma_8$ values. For this, we first use Eq.~\eqref{Eq: a - one param PA} to calculate $A_n$ for given ($\Omega_m$, $\sigma_8$), and then infer $n(>\nu)$ using Eq.~\eqref{eq:PA power law b(a)}. The accuracy of this prediction is shown by circle symbols in the right panel of Fig.~\ref{fig:PA cosmo and stats}. We quantify the success of the method in terms of the mean percentage residual, which is defined as the absolute value of the fractional difference between the measured peak abundance and the predicted one, averaged over all bins in $\nu\in[1.5,3.5]$. We find a mean percentage residual of $1-5$ percent, indicating that the model performs well. To understand what is the major factor affecting the model accuracy, we also calculate the mean percentage residual between the measured $n(>\nu)$ and the direct one-parameter power-law fit to it, which is shown as triangles in the right panel of Fig.~\ref{fig:PA cosmo and stats}. There is only a very slight difference between the triangles and the circles, with the former being generally lower. This indicates that our prediction of the peak abundance is very similar in accuracy to the original fit, which is further supported by the fact that the blue line in the left panel fits the various $A_n$ data points rather well. In summary, our model is able to predict the peak abundance to within ${\sim}3$ percent accuracy for most of the chosen cosmological models.

\begin{figure*}
    \centering
    \includegraphics[width=1.\textwidth]{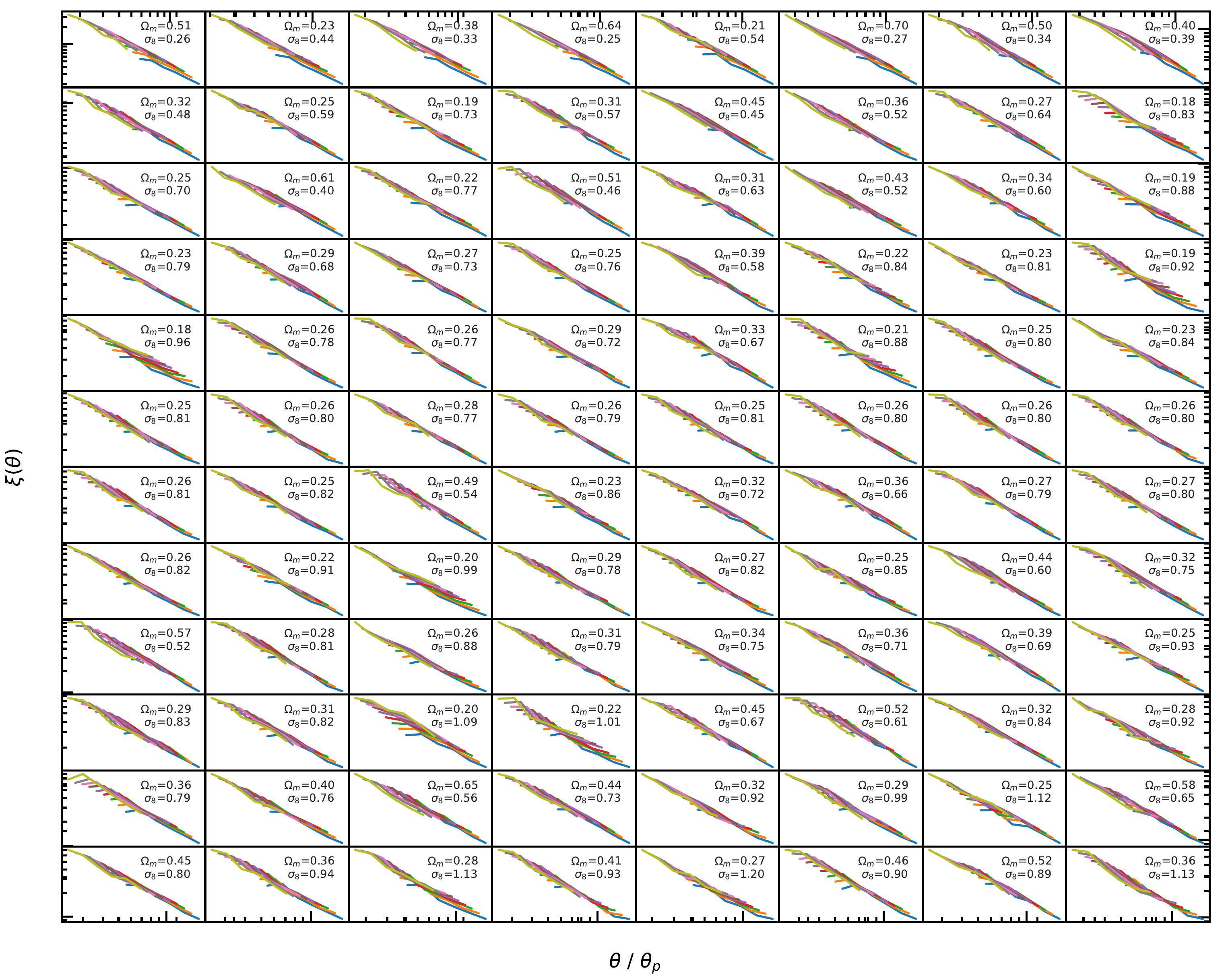}
    \vskip -.3cm
    \caption{The rescaled 2PCFs as a function of $\theta/\theta_p$ for various cosmological models. Each panel corresponds to a pair of $(\Omega_m,\sigma_8)$ parameters (see labels in each panel). The lines in each panel correspond to peak catalogues with different $\nucut$ thresholds, with $\nucut$ varying from $1.5$ to $3.5$ in $\Delta \nucut = 0.25$ increments. We find that all cosmologies have self-similar 2PCFs for peak catalogues with $\nucut\in[1.5,3.5]$. The x- and y-axis amplitudes of each sub-panel have been normalised to their respective centers to highlight the presence of the rescaled self-similarity across all of the $\Omega_m$, $\sigma_8$ models.}
    \label{fig:matrix plot}
\end{figure*}

If we use the original form of the power law, Eq.~\eqref{eq:PA power law1}, to model the peak abundance, with the two parameters, $A_n$ and $B_n$, both left to vary freely in the fitting, we get fits and predictions that match the raw data at the sub percent level. However, in attempt to minimise the number of parameters in our model we chose the one parameter power law, Eq. \eqref{eq:PA power law}, at the cost of roughly a $2\%$ loss in accuracy.

\subsection{Peak two-point correlation functions}
\label{section: rescaled 2PCF model}

\begin{figure}
    \centering
    \includegraphics[width=\columnwidth]{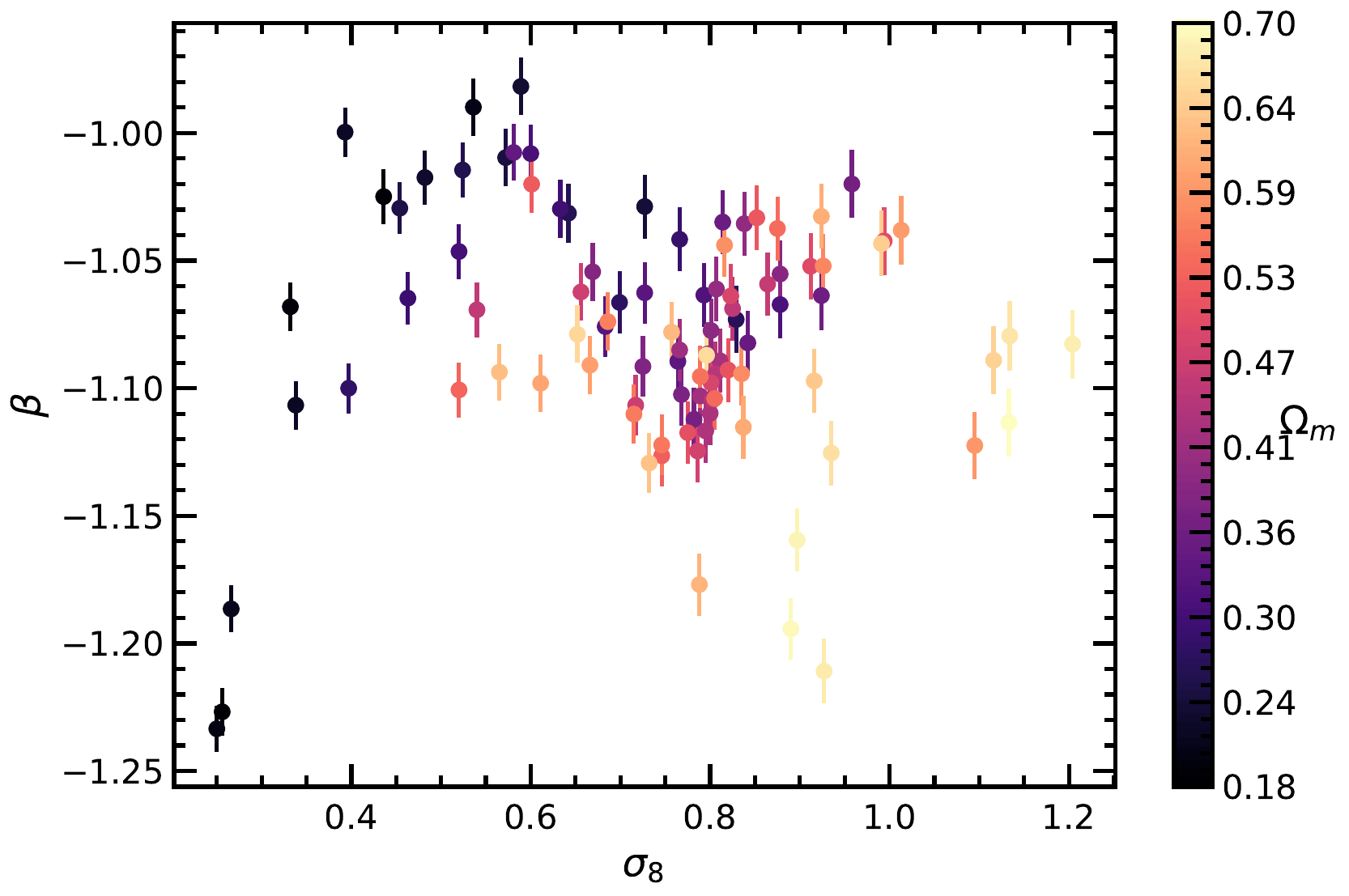}
    \vskip -.3cm
    \caption{The gradient, $\beta$, of the power laws fitted to each of the rescaled self-similar 2PCFs in Fig. \ref{fig:matrix plot} plotted against the $\sigma_8$ of the respective model with the associated $\Omega_m$ value given by the colour-bar. The vertical bars show the uncertainties in determining $\beta$. }
    \label{fig:beta}
\end{figure}

We now move on to check whether peak 2PCFs display the self-similar behaviour described in Section \ref{section: 2PCF rescaling and self-similarity} for a wide range of cosmological models. The result is shown in Fig.~\ref{fig:matrix plot}, where we plot the rescaled 2PCFs for all the pairs of $(\Omega_m,\sigma_8)$ values of the \citetalias{Matilla2017} maps (96 models in total). The re-scaled 2PCFs are normalised to the center of the panels to exemplify the self-similar behaviour. Fig.~\ref{fig:matrix plot} illustrates that the self-similarity of the 2PCFs is indeed robust against the change of cosmological parameters $\sigma_8$ and $\Omega_m$. The parameter space in Figs.~\ref{fig:cosmo models} and \ref{fig:matrix plot} is much larger than what is allowed by current constraints. Thus a model describing this self similarity will not only have the potential to provide additional cosmological constraints, but can also be applied to scenarios where predictions for a large parameter space is required, such as generating training sets for machine learning algorithms. 

Fig.~\ref{fig:matrix plot} also shows that the rescaled 2PCFs are well described by power laws which have very similar slopes across all cosmological models. To be more quantitative, we have fitted the following power-law function,
\begin{equation}
    \xi = \xi_0 \bigg( \frac{\theta}{\theta_p} \bigg) ^{\beta} \, ,
    \label{eq: rescaled 2pcf def}
\end{equation}
to the curves in each of the panels, with each data point weighted by its standard error (see Appendix \ref{Appendix: error estimation} for details about the error calculation). The best-fitting slope, $\beta$, as a function of cosmological parameters is shown in Fig.~\ref{fig:beta}, where the horizontal axis shows $\sigma_8$ while the colour of the points indicates the $\Omega_m$ value. The scatter of the points in Fig.~\ref{fig:beta} does not follow any clear trends, and combinations of $\sigma_8$ and $\Omega_m$ in the form of $\sigma_8 (\Omega_m / 0.3)^{\alpha}$, where $\alpha$ is allowed to vary, does not to lead to improvements in the correlation between $\beta$ and the input cosmology. In Fig. \ref{fig:beta} we find that the mean value across the entire sample is $\beta\approx-1.1$. As such, for simplicity, we take $\beta = -1.1$ as the power-law slope of the rescaled 2PCF (we have checked that the results are not particularly sensitive to the value of $\beta$).

\begin{figure*}
    \centering
    \includegraphics[width=2\columnwidth]{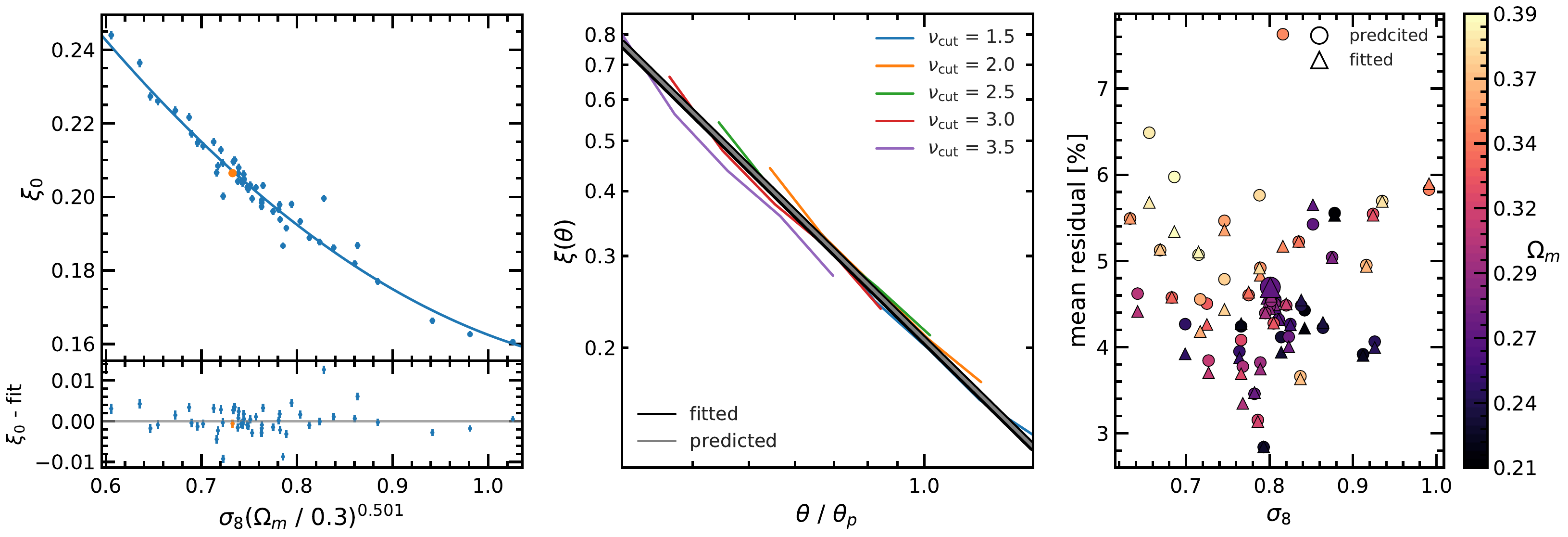}
    \vskip -.3cm
    \caption{\textit{Left panel:} the amplitude parameter of the power-law best fitting the rescaled 2PCFs as a function of $\sigma_8$ and $\Omega_m$. Each blue point shows the amplitude for a pair of $(\sigma_8,\Omega_m)$ values and the solid line shows the best fitting quadratic function.
    \textit{Middle panel:} an example of the rescaled 2PCF for different $\nucut$ values and and its best fitting power law (black line). The solid grey line shows our reconstructed power law, which was calculated using the best fitting line from the left panel.
    \textit{Right panel:} the percentage residuals between the fitted power law and the data (triangles), and between the reconstructed power law and the data (circles). The x-axis ($\sigma_8$) and the colour-bar ($\Omega_m$) indicate the cosmology of the model for which the residuals are being measured}
    \label{fig:selfsim cosmoexamplestats}
\end{figure*}

On the other hand, we find that the amplitude of the rescaled peak 2PCFs, $\xi_0 = \xi_0(\Omega_m, \sigma_8)$ in Eq.~\eqref{eq: rescaled 2pcf def}, shows a systematic dependence on the cosmological parameters. Therefore, in order to have a complete description we also need to model $\xi_0(\Omega_m, \sigma_8)$.
The fitting result for $\xi_0$ for the selected cosmological models (orange points in Fig.~\ref{fig:cosmo models}) is displayed in the left panel in Fig.~\ref{fig:selfsim cosmoexamplestats}, where it is plotted against $\sigma_8 (\Omega_m/0.3)^{0.501}$ with the index $\alpha=0.501$ characterising the degeneracy direction between $\Omega_m$ and $\sigma_8$ for the rescaled 2PCF amplitude $\xi_0$. The value $0.501$ is tuned such that the data points on the left panel of Fig.~\ref{fig:selfsim cosmoexamplestats} are fitted using a smooth quadratic curve with the lowest $\chi^2$. This is shown as the blue solid line in the left panel of Fig.~\ref{fig:selfsim cosmoexamplestats},  
which takes the form 
\begin{equation}
    \xi_0  = \xi_{0,a}x^2 + \xi_{0,b}x + \xi_{0,c} \, ,
    \label{eq: amplitude cosmo quadratic}
\end{equation}
where
\begin{align}
    \xi_{0,a} =  0.253 \, , \;\;\;\;\;
    \xi_{0,b} = -0.605 \, , \;\;\;\;\;
    \xi_{0,c} =  0.514 \, ,
    \label{xi_0 quad params}
\end{align}
and $x = \sigma_8 (\Omega_m / 0.3)^{0.501}$. The lower sub-panel of the left panel shows the residual between the measured amplitude $\xi_0$ and its fitted values. The residuals show no systematic trends with varying $\sigma_8(\Omega_m/0.3)^{0.501}$, indicating that the fitting function works equally well for all cosmologies. 

In the middle panel of Fig.~\ref{fig:selfsim cosmoexamplestats} we have randomly selected one of the cosmologies from the \citetalias{Matilla2017} maps, and compared the rescaled 2PCFs at several $\nucut$ values between $1.5$ and $3.5$ (coloured lines), the power-law fit to these rescaled 2PCFs (black solid line, which we call the `fitted' curve), and the predicted rescaled 2PCF for this particular cosmology (the grey straight line, which we call the `prediced' curve). The latter was obtained by calculating $\xi_0$ using Eq.~\eqref{eq: amplitude cosmo quadratic}, and then inferring the 2PCF from Eq. \eqref{eq: rescaled 2pcf def}.  This matches the original fitted power law very closely, indicating that the model described by Eqs.~(\ref{eq: rescaled 2pcf def}) and (\ref{eq: amplitude cosmo quadratic}) works very well.

We next quantify the accuracy of our prediction for the rescaled 2PCF. For a given cosmological model, such as the one shown in the middle panel of Fig.~\ref{fig:selfsim cosmoexamplestats}, we calculate the residuals, i.e., the fractional differences of the `fitted' and `reconstructed' curves with respect to the rescaled measured 2PCFs. This is done for each of the five $\nucut$ values shown in Fig.~\ref{fig:selfsim cosmoexamplestats}, and we define the mean residual as the average over all $\theta/\theta_p$ bins and all $\nucut$ values. The mean residuals for the fitted and predicted curves are respectively shown by a large triangle and a large circle in the right panel of Fig.~\ref{fig:selfsim cosmoexamplestats}. We have repeated this procedure for all the cosmological models and have plotted their residuals in the right panel, with the associated $\sigma_8$ values shown in the x-axis and $\Omega_m$ values shown by the colour bar to the right. We find that the model prediction is almost as accurate as the direct fitting, and is able to match the rescaled 2PCFs at about 5\% accuracy level. The large symbols in the right panel correspond to the model shown in the middle panel, to give a visual illustration about how well the 2PCF model in Eqs.~(\ref{eq: rescaled 2pcf def}, \ref{eq: amplitude cosmo quadratic}) works for an `average' cosmology for which the mean residual is $4.8\%$.

\begin{figure}
    \centering
    \includegraphics[width=\columnwidth]{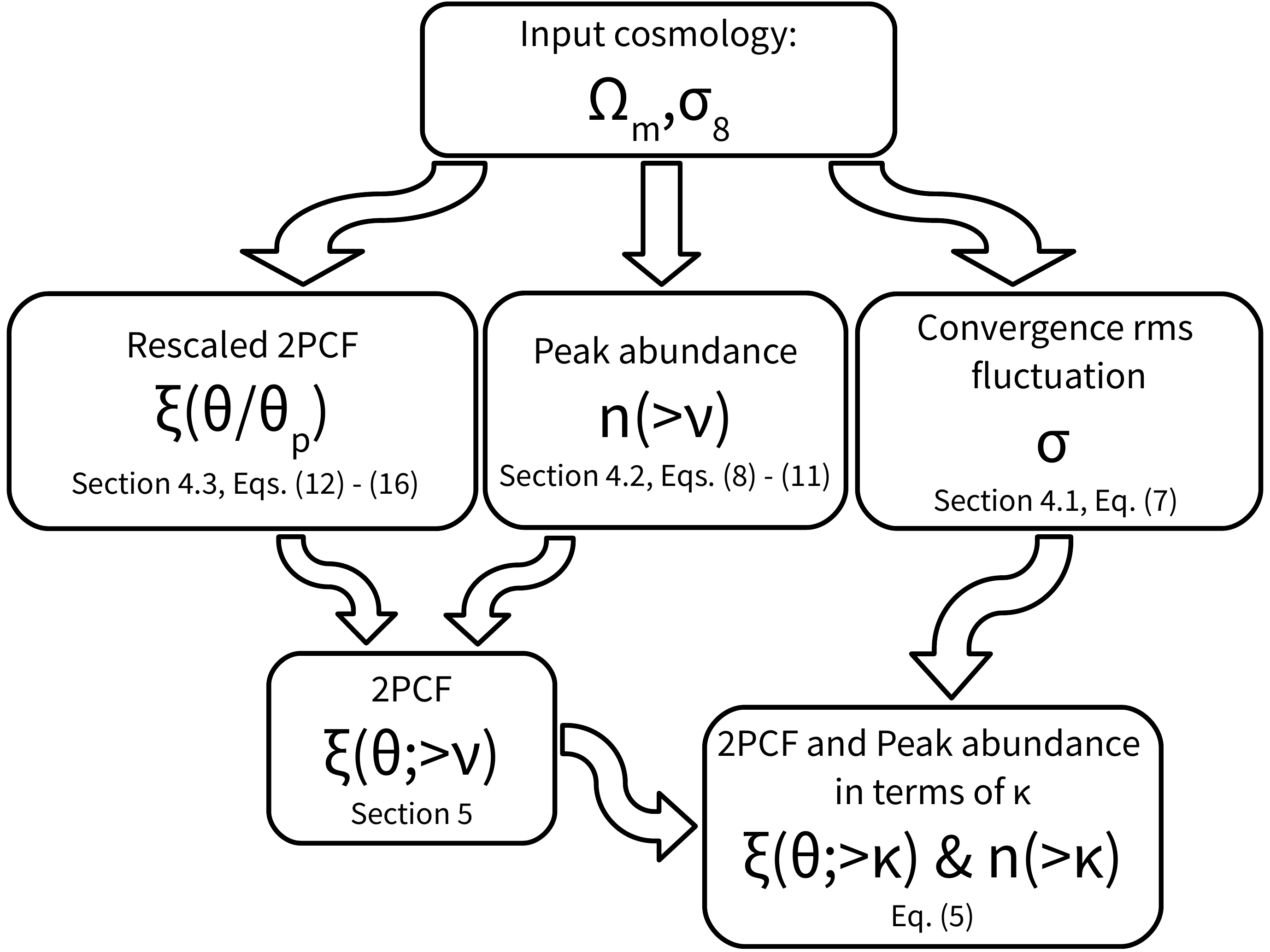}
    \caption{This flowchart describes the pipeline our model uses to reconstruct the peak 2PCF by exploiting its self similarity. First we take input cosmological parameters, $\Omega_m$ and $\sigma_8$, which our model uses to predict the rescaled 2PCF, the peak abundance and the rms fluctuations of the convergence map. These statistics can then be combined to give the original 2PCF for peaks of different heights, expressed in terms of either $\nu$ or $\kappa$.
    }
    \label{fig:flowchart}
\end{figure}

\section{A pipeline for 2PCF Reconstruction}\label{section: reconstruction pipeline}

\begin{figure*}
    \centering
    \includegraphics[width=\textwidth]{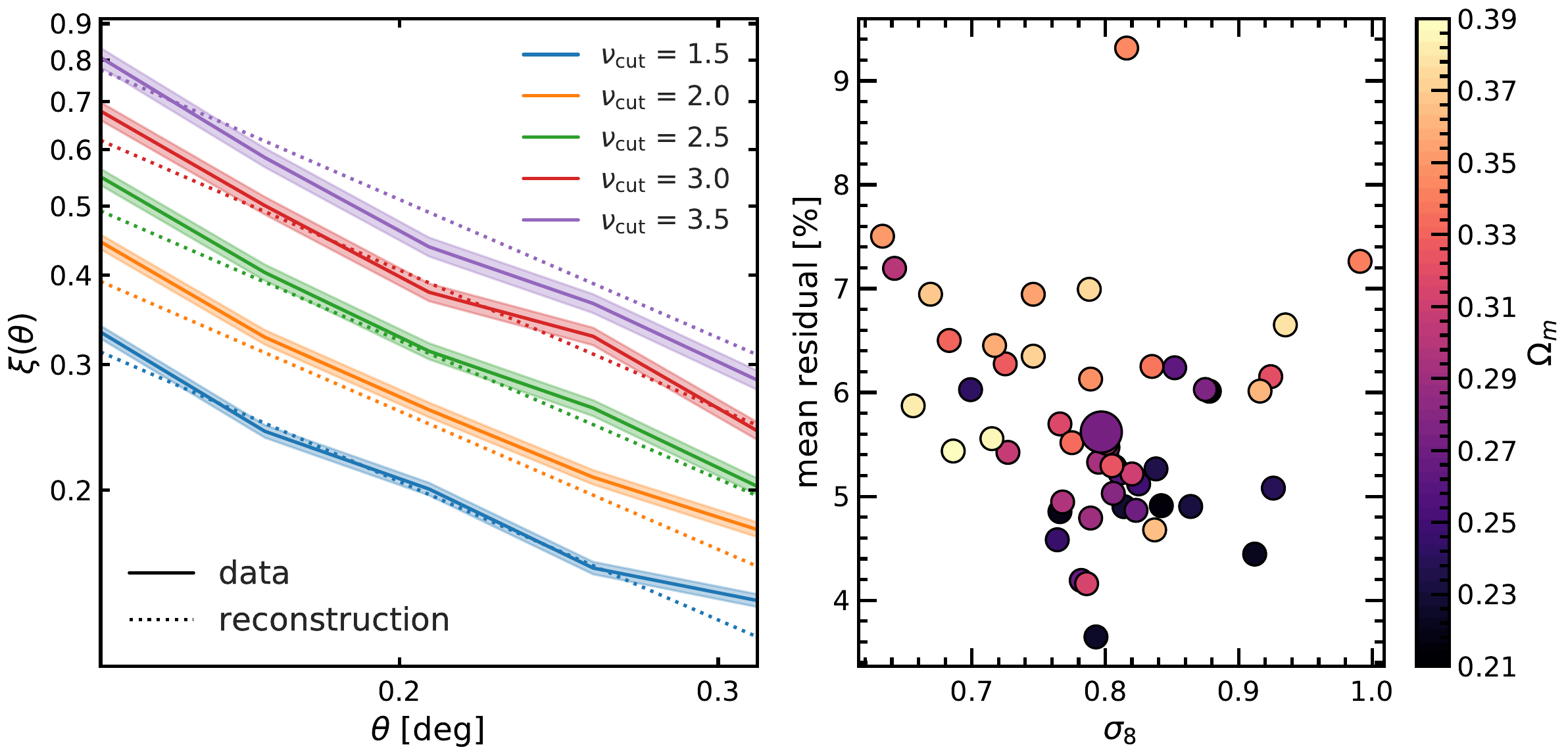}
    \vskip -.3cm
    \caption{ \textit{Left panel:} Reconstructed 2PCFs from our model (dashed) compared to measured 2PCFs from N-body simulations (solid), for peak catalogues with $\nu_{\rm{cut}} \in [1.5,3.5]$.
    \textit{Right panel:} Mean percentage residuals between the reconstructed and measured 2PCFs. The larger symbol indicates the example model that is shown in the left panel. The x-axis and colour-bar indicate the $\sigma_8$ and $\Omega_m$ values of the models respectively.}
    \label{fig:reconstruction}
\end{figure*}

We can combine the models developed in the previous section for the convergence rms fluctuation, peak abundance and rescaled peak 2PCF, to develop an integrated pipeline that allows us to predict the (un-rescaled) peak 2PCFs, $\xi(\theta)$, as a function of $\nucut$. The procedure is schematically illustrated in Fig.~\ref{fig:flowchart} and outlined as follows: 
\begin{enumerate}
    \item For chosen $\Omega_m$ and $\sigma_8$ values, one can use the models to predict the peak abundance (Section \ref{section: Peak abundance model}) and the rescaled 2PCF (Section \ref{section: rescaled 2PCF model}).
    \item These two statistics are combined, using $\theta_p = 1/\sqrt{n(> \nu_{\rm{cut}})}$, to give the 2PCF, $\xi(\theta)$, for peak catalogues with $\nu_{\rm{cut}} \in [1.5,3.5]$.
    \item If needed, the above-predicted peak abundance and 2PCFs can then be expressed in terms of $\kappa$ by using the $\sigma(\Sigma_8)$ fit in Section \ref{section: Convergence map standard deviation}.
\end{enumerate}  
This pipeline offers a simple apparatus to make predictions of the one- and two-point statistics for intermediate ($\nu$ $\in$ [1.5,3.5]) WL peaks, which can be used (on its own or together with other cosmological probes) to constrain the parameters $(\sigma_8, \Omega_m)$ using observational data. It will be interesting to see if these new statistics are complimentary to other probes, such as the shear-shear correlation, when constraining ($\Omega_m$,$\sigma_8$), but this will be left for future follow up works. In the next section we will discuss further aspects which need to be checked before applying this method.

As a proof of concept, we show an example of this 2PCF reconstruction pipeline in the left panel of Fig.~\ref{fig:reconstruction}. The solid curves show the 2PCFs measured from the simulation data for an arbitrarily selected cosmology, with shaded regions showing the (under) estimated standard error (see Appendix \ref{Appendix: error estimation} for more detail). The dashed lines show the predictions by our 2PCF reconstruction pipeline. We find a reasonably good agreement between the simulated and reconstructed 2PCFs, with the latter mostly lying within or just outside the (under)estimated errors bars. The second panel in Fig.~\ref{fig:reconstruction} shows the mean percentage difference between the reconstructed and measured 2PCFs, averaged over the 5 plotted 2PCFs and all $\theta$ bins with $\nu_{\rm{cut}} \in [1.5,3.5]$ separated by a $\Delta\nu_{\rm{cut}} = 0.5$ increment. The model that has been selected to exemplify the reconstruction is indicated by the large symbol in the left panel of Fig. \ref{fig:reconstruction}, which is an "average" one in terms of the performance the reconstruction (there are many models for which the reconstruction works better). We can see that for all of the selected cosmologies, our model is able to predict the 2PCF to within a roughly $6\%$ uncertainty on average. Relative to the estimated errors bars, the quality of our reconstruction is reasonably good. 

We find that generally the amplitude of the 2PCFs is overestimated for the larger $\nu_{\rm{cut}}$ catalogues. This could be a fundamental aspect of the 2PCF evolution, however due to small map sizes and low peak number densities (at approximately 7 deg$^{-2}$) it is likely that 2PCFs with $\nu > 3.5$ are biased. The true amplitude of the 2PCFs with larger $\nu_{\rm{cut}}$ could be measured more accurately with larger weak lensing maps, which we leave to further study.

\section{The impact of galaxy shape noise}\label{section:GSN}

\begin{figure}
    \centering
    \includegraphics[width=\columnwidth]{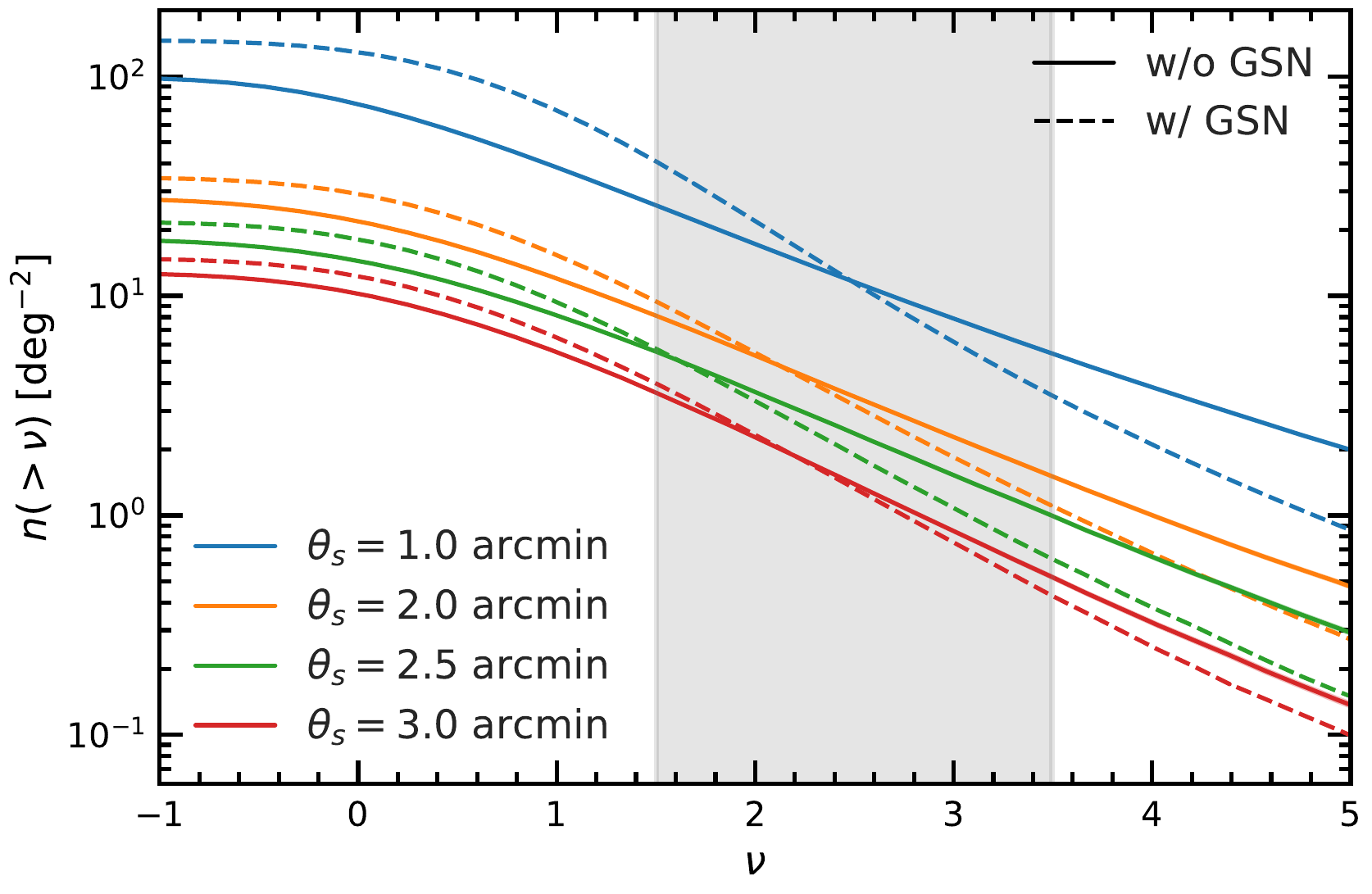}
    \caption{Weak lensing peak abundances for four smoothing scales, $\theta_s$ = 1.0, 2.0, 2.5 and $3.0\arcmin$ (see labels), for peaks extracted from convergence maps without GSN (solid) and for peaks extracted from convergence maps with added GSN (dashed). Here the added GSN matches LSST specifications ($\sigma_{\rm{int}} = 0.3$, $n_{\rm{gal}} = 40$ arcmin$^{-2}$).
    }
    \label{fig:PA_GSN}
\end{figure}

\begin{figure*}
    \includegraphics[width=.98\textwidth]{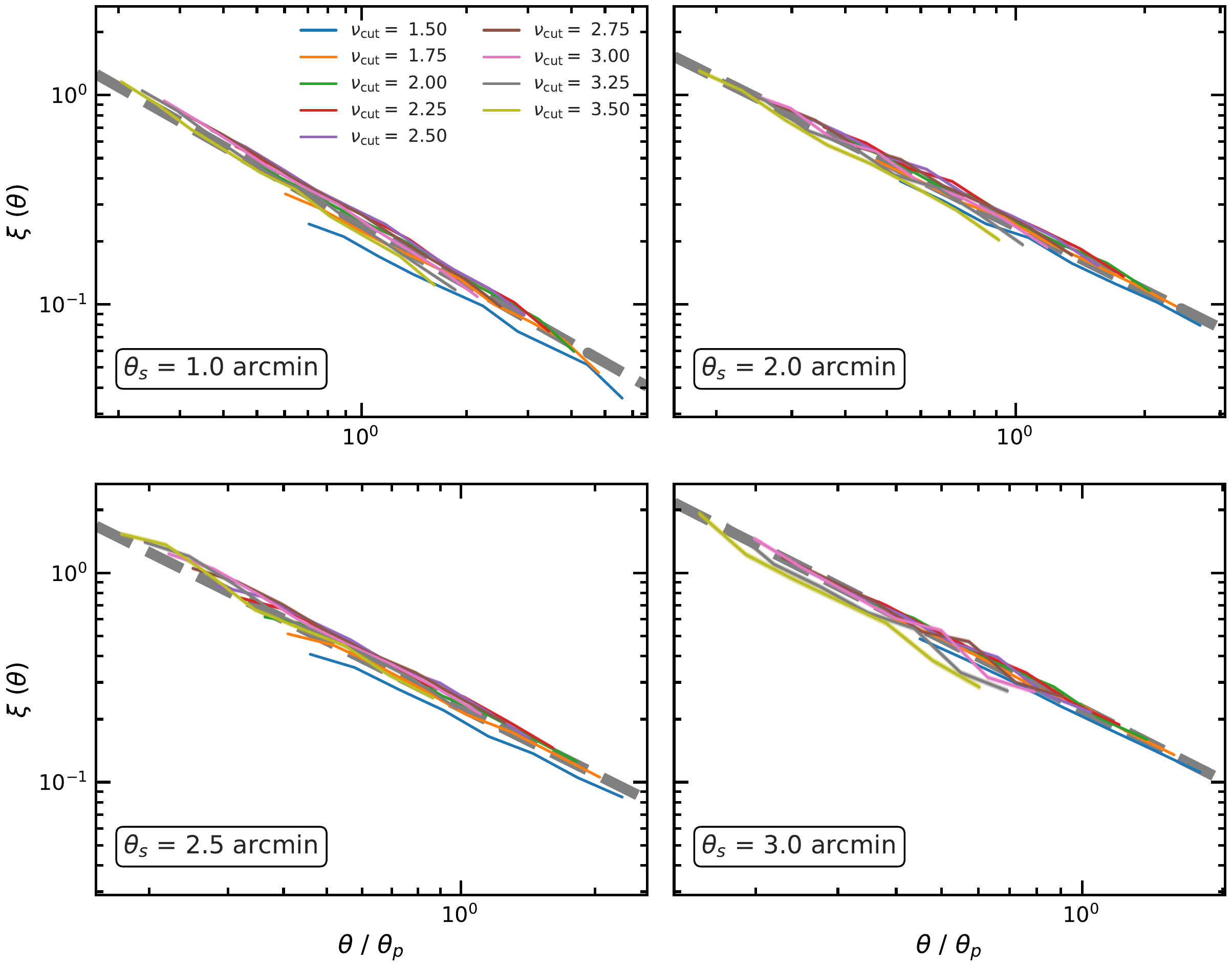}
    \vskip -.3cm
    \caption{Rescaled 2PCFs for four smoothing scales, $\theta_s$ = 1.0, 2.0, 2.5 and $3.0\arcmin$ (see label in each panel), for peak catalogues extracted from convergence maps with added GSN that matches LSST specifications ($\sigma_{\rm{int}} = 0.3$, $n_{\rm{gal}} = 40$ arcmin$^{-2}$). The various solid coloured lines correspond to peak catalogues with different $\nucut$ thresholds (see legend in the upper left panel), with $\nucut \in [1.5,3.5]$ incremented in steps $\Delta \nucut = 0.25$. The grey thick dashed lines show the fits to the rescaled 2PCFs for the same smoothing scales but without GSN.}
    \label{fig:2PCF GSN}
\end{figure*}

Up to here, we have discussed the WL peak abundance and 2PCF in a theoretical context with the aim of having a model that allows us to accurately describe and predict these statistics in an idealised situation. Whilst this theoretical model can have useful applications in, e.g., mock WL peak catalogue generation, to be more useful for cosmological constraints, we need to investigate the self similarity of the 2PCF in more realistic situations. One of the things we have not included in our analysis so far is galaxy shape noise (GSN). 

GSN is a source of uncertainty in WL observations, where the measured ellipticity of galaxies is dominated by their random orientation, and only weakly correlated due to gravitational lensing on scales much larger than the galaxy-galaxy separation. Observations of cosmic shear, and therefore cosmic convergence, is contaminated by this noise. One usually uses large smoothing lengths to suppress this noise in order to recover statistics more reliably. However, large smoothing lengths could either dampen the amplitude of the measured statistics, which is evident from the decrease of the WL peak abundance with increasing smoothing scales in Fig.~\ref{fig:peaks abundance}, or increase the noise in the measurements, which can be seen to a small extent in the 2PCFs for different smoothing lengths in Fig.~\ref{fig:self sim 1and2}. Therefore, a trade-off has to be struck between using a large enough smoothing length in order to suppress the galaxy shape noise, and not over smoothing so that interesting statistics are not suppressed more than they need to be. With convergence maps from N-body simulations, we can test the difference in the peak abundance and 2PCF for cases with and without GSN for a range of smoothing lengths. 

For this section, we include GSN in the \citetalias{Takahashi2017} convergence maps that match LSST specifications by adding to each pixel within a map random values drawn from a Gaussian distribution with a standard deviation $\sigma_{\rm{pix}}$ given by
\begin{equation}
    \sigma_{\rm{pix}}^2 = \frac{ \sigma_{\rm{int}}^2 }{ 2 \theta_{\rm{pix}}^2 n_{\rm{gal}} } \, ,
    \label{GSN spread}
\end{equation}
where $\sigma_{\rm{int}}$ is the dispersion of the intrinsic source galaxy ellipticity, $\theta_{\rm{pix}}$ is the angular width of the pixel to which noise is added and $n_{\rm{gal}}$ is the number density of source galaxies. To match LSST specification we use $\sigma_{\rm{int}} = 0.4$ and $n_{\rm{gal}} = 40$ arcmin$^{-2}$ \citep{LSST2009}

After GSN is added to the pixels, we smooth the maps, identify peaks in the noise-added smoothed WL maps using Eq.~(\ref{Eq:nu}) with $\sigma$ also directly measured from the noisy maps, recalculate the peak abundance and peak 2PCFs, and compare these statistics to the case with no GSN, with the same smoothing.

The impact of GSN on the WL peak abundance is shown in Fig.~\ref{fig:PA_GSN}, where the solid and dashed lines respectively correspond to peaks identified in WL maps with and without GSN. Here we study four smoothing scales, $\theta_s=1$, $2$, $2.5$ and $3$ arcmin. In each instance, $\nu$ is defined relative to the WL map in which the peaks are identified, so for the GSN added case, $\sigma$ in Eq.~\eqref{Eq:nu} includes contributions to the rms fluctuations from both GSN and the underlying convergence signal, while for the no GSN case $\nu$ is defined by taking $\sigma$ as the rms convergence fluctuation.

For all smoothing lengths, by adding GSN, the peak abundance increases at low $\nu$ and decreases at high $\nu$, with a crossover between $\nu = 1.5$ and $2.5$ depending on the smoothing scale. GSN has the largest impact on the peak abundance for the smallest smoothing length, while for larger $\theta_s$ the agreement between the peak abundances in the GSN and no GSN cases is better, although substantial difference remains even in the case of $\theta_s=3$ arcmin. This means that the fitting formulae, Eqs.~(\ref{eq:PA power law}, \ref{eq:PA power law b(a)}), which describe the cosmology dependence of peak abundance, need to be recalibrated by using peaks extracted from GSN-added maps. Due to the small size of the \citetalias{Matilla2017} WL maps, this will be left as future work when larger simulations of different cosmologies are available.

Note that in Fig.~\ref{fig:PA_GSN} the peaks are defined using Eq.~(\ref{Eq:nu}), where $\sigma$ is the total rms convergence that includes contributions from the physical rms convergence and from the rms of noise. This explains the crossover mentioned above: because $\sigma$ is increased, for the high peaks their $\nu$ values actually decrease, and the number of such high peaks does not increase quickly enough to maintain $n(<\nu)$ at large $\nu$, which causes the latter to drop compared with the no GSN case. We have explicitly checked (not shown here) that, if one defines $\nu$ in Eq.~(\ref{Eq:nu}) by using the same $\sigma$ for the GSN and no GSN cases, then the peak abundance is consistently higher in the former case, due to artificial peaks created by noise.

In order to closely inspect the impact of using different smoothing lengths on the self similarity of the rescaled 2PCFs, we have tried four different $\theta_s$ values, respectively $\theta_s=1$, $2$, $2.5$ and $3$ arcmin. The results are shown in Fig.~\ref{fig:2PCF GSN}, where the peaks are all identified from, and the $\sigma$ used to define the SNR $\nu$ are all measured by using, the smoothed noisy maps. 
Interestingly, we find that the rescaled 2PCFs are still on top of each other for all four smoothing lengths. With $\theta_s=1\arcmin$, the agreement between the rescaled 2PCFs is weaker, where only the curves with $2 \leq \nu_{\rm{cut}} \leq 3.5$ appear to be self similar. For $2 \arcmin{}$ smoothing the 2PCFs appear to be self similar in the entire $1.5 \leq \nu_{\rm{cut}} \leq 3.5$ range, and shows that the self-similarity of 2PCFs is robust against GSN. With $2.5 \arcmin{}$ smoothing the overall self similarity appears to be tighter, however the $\nu_{\rm{cut}} = 1.5$ appears to be outside the self similar range.  Finally, for $3 \arcmin{}$ smoothing we see that the self similarity of the 2PCFs holds up to $\nu_{\rm{cut}} =3$, after which the rescaled 2PCFs drop off in amplitude. It is possible that this drop in amplitude is caused by the small map size (\map{10}) and low number density of tracers ($\approx$ 0.5 deg$^{-2}$), rather than a breakdown of the self similarity. As $\theta_s$ increase it also appears that the overall gradient of the rescaled 2PCFs decreases.

Having verified that the 2PCFs remain self similar in the presence of GSN, next we want to see how including the latter affects the power law of the rescaled self-similar peak 2PCFs. In each subpanel of Fig.~\ref{fig:2PCF GSN}, we have overplotted, as the grey dashed lines, the best-fit power-law functions for the rescaled 2PCFs of the peaks extracted from the \citetalias{Takahashi2017} maps smoothed using the same $\theta_s$ values but without adding GSN (the grey dashed lines in the top two panels of Fig.~\ref{fig:2PCF GSN} are the same as the black dashed lines in Fig.~\ref{fig:self sim 1and2}). The two cases are in good agreement for all four smoothing scales, which shows that the impact of GSN on the rescaled 2PCF is minor. This is a nice property, since it indicates that GSN will not significantly contaminate the underlying cosmology dependence of the rescaled 2PCF if the same observation applies to other cosmologies. However, due to the limited map size from \citetalias{Matilla2017} we leave this investigation to future study. 

In short, we conclude that the prevalence of the self similarity in the 2PCFs for peaks extracted from GSN-added WL maps shows that this feature is robust to this observational systematic, and therefore has the potential to be used in cosmological constraints.

\section{Discussion and Conclusions}
\label{section: Discussion}

In this paper, we have investigated the one- and two-point statistics for intermediate peaks, with SNR values $\nu\in[1.5,3.5]$, from weak lensing convergence maps. These WL peaks contain useful information about the LSS formation, and the analyses of them are expected to place complementary constraints on the cosmological model. However, unlike high peaks, the intermediate WL peaks are not individually associated to the most massive dark matter structures, making the modelling of their statistical properties more challenging. To overcome this difficulty, we rely on WL convergence maps constructed from a large number of N-body simulations with varying cosmological parameters and technical specifications, to attempt to find patterns of the peak statistics and their cosmology dependence. Our main findings are summarised as follows:
\begin{itemize}
    \item The rms fluctuation of WL convergence, $\sigma$, has a linear dependence on a particular combination of $\Omega_m$ and $\sigma_8$ via $\sigma_8\left(\Omega_m/0.3\right)^\alpha$, with the parameter $\alpha$ weakly dependent on the smoothing length of the convergence map, cf.~Fig.~\ref{fig:std cosmo}. This linear dependence is given in Eq.~\eqref{Eq: sigma relation}, and highlights a universal behaviour within $\Lambda$CDM which may be exploited to make cosmological constraints.

    \item A universal one-parameter power law function is found, which can describe the WL peak count for $\nu\in[1.5,3.5]$ with an accuracy of within $\approx1$-$5\%$, for a large range of $\Omega_m$ and $\sigma_8$ values, cf.~Fig.~\ref{fig:peaks abundance} and Eq.~\eqref{eq:PA power law}. The accuracy of the power-law description of the peak abundance can reach the sub-percent level if two free parameters are used in the power-law function.
    
    \item A self-similar behaviour of the WL peak 2PCF has been found by rescaling the angular separation, $\theta$, between a pair of peaks by the mean inter-peak separation, $\theta_p$. While the amplitude of the original 2PCF increases with $\nucut$, the rescaled 2PCFs for $\nucut\in[1.5,3.5]$ lie on top of each other cf.~Fig.~\ref{fig:self sim 1and2}.
    
    \item This self-similar behaviour holds for a very wide range of $(\Omega_m, \sigma_8)$ values, and we find a simple quadratic dependence of the amplitude of the rescaled 2PCFs on $\sigma_8\left(\Omega_m/0.3\right)^\alpha$, while the slope of the rescaled 2PCFs have negligible dependence on $\Omega_m$ and $\sigma_8$, cf.~Figs.~\ref{fig:matrix plot}, \ref{fig:beta} and \ref{fig:selfsim cosmoexamplestats}. A fitted model to predict the peak 2PCF for any chosen $\Omega_m$ and $\sigma_8$ is given in Eq.~\eqref{eq: rescaled 2pcf def}.

    \item A pipeline is developed which combines the above three fitted models, for the convergence rms fluctuation, WL peak abundance and rescaled peak 2PCF respectively, to predict the raw peak 2PCF $\xi(\theta;\nucut)$ for $\nucut\in[1.5,3.5]$ and any given $\Omega_m$ and $\sigma_8$ with good accuracy, cf.~Fig.~\ref{fig:reconstruction}. 
    
    \item We found that the self similarity of the peak 2PCF holds in the presence of galaxy shape noise and larger smoothing lengths, cf.~Fig.~\ref{fig:2PCF GSN}.
\end{itemize}

The most important application of the results presented in this work is in constraining the $\Omega_m$ and $\sigma_8$ cosmological parameters. As demonstrated above, the pipeline integrating the models for WL peak abundance and self-similar rescaled 2PCFs is able to reconstruct the raw, unrescaled, peak 2PCFs for various $\nucut$ values with a typical accuracy of better than $6\%$. Furthermore, we have seen that the WL peak abundance and 2PCFs depend on very different combinations of $\Omega_m$ and $\sigma_8$, one with $\sigma_8\left(\Omega_m/0.3\right)^{-0.638}$ and the other $\sigma_8\left(\Omega_m/0.3\right)^{0.501}$. This indicates that a simultaneous use of these statistics already holds the potential of breaking the degeneracy between $\Omega_m$ and $\sigma_8$ before including other cosmological probes. 
\cite{Marian2013} found that the 2PCFs of high WL peaks only provide weakly complimentary constraints on ($\Omega_m$,$\sigma_8$) when combined with the peak abundance. In this work we investigate the 2PCFs of WL peaks with intermediate heights and above, as well as combining the 2PCFs from multiple peak catalogues in the form of a rescaled 2PCF described by a single power law. The powerlaw describing the rescaled 2PCF may be more sensitive to cosmology than the 2PCF of high peaks.

We note that the degeneracy direction of the peak abundance of intermediate height peaks, which are studied in this paper, are very different to that of low and high peaks, which has also been observed in \cite{J.Liu2015} and explained in \cite{Yang2011}. Therefore, using the counts of intermediate height peaks may be complimentary to using the full peak abundance and could aid in breaking the $\Omega_m$ and $\sigma_8$ degeneracy.

Another potential application of our results is the generation of mock WL peak catalogues. For a given input cosmological model, the pipeline can be used to predict the WL peak counts and 2PCFs as described above. Random realisations of peaks can then be generated with the peaks arranged such that they have the desired number density and spatial clustering. One technique to do this is point process \citep[see, e.g.,][for some recent progress and applications]{Oztireli2012}. This is a Monte Carlo approach where a candidate point (e.g., a WL peak) is placed in a field, which is accepted if its inclusion into the field pushes the measured 2PCF closer to the input one, and rejected otherwise. Point process is a well-developed and widely-used technique to generate point catalogues. In the WL peak case, the situation is slightly more complicated, because the generated catalogue should have peaks of different SNR (or $\nu$ values), which simultaneously have the desired 2PCFs at different $\nucut$ values. We expect that the good agreement between the rescaled peak 2PCFs will prove useful in dealing with this issue, though a detailed investigation into this interesting question will be left for a future work. The fast generation of mock WL peak catalogues can be used for evaluating covariance matrices and studying other cosmological quantities, such as voids identified from WL peaks \citep{Davies2018}.

The proof-of-concept study in this work has also left various possible further extensions of the analyses presented here. One of the most important considerations for future WL surveys and their cosmological applications is the effect of galaxy shape noise. Using the all-sky maps from \citetalias{Takahashi2017}, we have shown that (i) the inclusion of GSN necessitates a larger smoothing length than used in the bulk of this paper, $\theta_s=2$-$3$ arcmin, to suppress its impact on the extracted cosmological statistics, and (ii) with a suitable smoothing, the self-similarity of the peak 2PCFs still holds for the cosmology used in the \citetalias{Takahashi2017} simulations. While we expect these conclusions to apply for other cosmological models, Fig.~\ref{fig:2PCF GSN} shows that the use of GSN and larger $\theta_s$ does indeed affect the slope of the rescaled peak 2PCF. Therefore, in the presence of GSN our fitted models need to be re-analysed before it can be directly useful for cosmological tests.

Unfortunately, the 96 \citetalias{Matilla2017} maps with varying cosmological parameters have a relatively small size, at $3.5\times3.5$ deg$^2$. Including GSN in these maps and increasing the smoothing length will reduce both the number of peaks in the maps and the dynamical range over which the 2PCFs can be reliably studied. This consideration makes a compelling case that larger convergence maps, constructed from N-body simulations with larger boxes and varying cosmologies, are a natural next step, to re-calibrate our peak models so that they can be readily applied for upcoming WL surveys. Again, we leave these to a future, more comprehensive, study.

The planned larger simulations will have other applications as well. For example, they will allow us to study low/intermediate WL peaks and the high peaks, as well as other statistics such as the WL shear power spectrum, simultaneously. It will also be possible to look at source galaxies with a certain redshift distribution compared to the currently idealised case with a single source redshift, $z_s=1$. Larger WL maps will also allow us to more accurately estimate the errors on the 2PCFs, with large-scale modes properly included. Further more, in future studies we will try methods of extracting WL peaks that are more similar to approaches taken in observations, such as starting with the shear field and adding GSN to this before we then transform to the convergence field. 

Finally, it will also be interesting to analyse the rescaled WL peak 2PCFs in cosmological models beyond $\Lambda$CDM. We can envisage two possible scenarios here: the first is that the rescaled 2PCFs may not be self similar, which would offer a potentially strong constraint on these models. Alternatively, the detailed properties of the self-similarity in the 2PCFs may change, in the form of a different amplitude or slope, which can also be used to test models with observational data. Therefore, it will be important to consider models which are expected to alter the large-scale clustering of matter. These include the various dark energy models which may couple to dark matter or have different equation-of-state $w$ parameters. The neutrino mass is another interesting possibility, as massive neutrinos tend to dampen structure formation, which leaves signatures in the WL peak abundance and 2PCF. Modified gravity models can also be potentially tested since they generally introduce fifth forces on cosmological scales, which modify the clustering of matter or even the geodesics of photons. The studies of these topics will require new simulations and dedicated effort, and will be deffered for the future.

\section*{Acknowledgements}
We thank Zuhui Fan, Daniel Gruen, Zoltan Haiman, Catherine Heymans, Chieh-An Lin, Jia Liu, Xiangkun Liu, Peder Norberg, Masato Shirasaki, for useful discussions and comments on this draft.

CTD is funded by a Science and Technology Facilities Council (STFC) PhD studentship through grant ST/R504725/1. MC acknowledges support by the European Research Council through an ERC Advanced Investigator Grant, DMIDAS [GA-786910]. BL is supported by an ERC Starting Grant, ERC-StG-PUNCA-716532. MC and BL are additionally supported by the STFC Consolidated Grants [Nos.~ST/I00162X/1, ST/P000541/1]

This work used the DiRAC facility, hosted by Durham University, managed by the Institute for
Computational Cosmology on behalf of the STFC DiRAC HPC Facility (www.dirac.ac.uk). The equipment was funded by BEIS capital funding
via STFC capital grants ST/K00042X/1, ST/P002293/1, ST/R002371/1 and ST/S002502/1, Durham University and STFC operations grant
ST/R000832/1. DiRAC is part of the National e-Infrastructure.

We thank the Columbia lensing group (http://columbialensing.org) for
making their suite of simulated lensing maps available, and the US
National Science Foundation (NSF) for supporting the creation of those
maps through grant AST-1210877 and XSEDE allocation AST-140041.


\bibliographystyle{mnras}
\bibliography{mybib}

\begin{thebibliography}{}
\makeatletter
\relax
\def\mn@urlcharsother{\let\do\@makeother \do\$\do\&\do\#\do\^\do\_\do\%\do\~}
\def\mn@doi{\begingroup\mn@urlcharsother \@ifnextchar [ {\mn@doi@}
  {\mn@doi@[]}}
\def\mn@doi@[#1]#2{\def\@tempa{#1}\ifx\@tempa\@empty \href
  {http://dx.doi.org/#2} {doi:#2}\else \href {http://dx.doi.org/#2} {#1}\fi
  \endgroup}
\def\mn@eprint#1#2{\mn@eprint@#1:#2::\@nil}
\def\mn@eprint@arXiv#1{\href {http://arxiv.org/abs/#1} {{\tt arXiv:#1}}}
\def\mn@eprint@dblp#1{\href {http://dblp.uni-trier.de/rec/bibtex/#1.xml}
  {dblp:#1}}
\def\mn@eprint@#1:#2:#3:#4\@nil{\def\@tempa {#1}\def\@tempb {#2}\def\@tempc
  {#3}\ifx \@tempc \@empty \let \@tempc \@tempb \let \@tempb \@tempa \fi \ifx
  \@tempb \@empty \def\@tempb {arXiv}\fi \@ifundefined
  {mn@eprint@\@tempb}{\@tempb:\@tempc}{\expandafter \expandafter \csname
  mn@eprint@\@tempb\endcsname \expandafter{\@tempc}}}

\bibitem[\protect\citeauthoryear{Albrecht et~al.}{Albrecht
  et~al.}{2006}]{Albrecht2006}
Albrecht A.,  et~al., 2006, preprint (\mn@eprint {arXiv} {astro-ph/0609591})

\bibitem[\protect\citeauthoryear{Amendola et~al.,}{Amendola
  et~al.}{2013}]{Amendola2013}
Amendola L.,  et~al., 2013, \mn@doi [LRR] {10.12942/lrr-2013-6}, 16, 6

\bibitem[\protect\citeauthoryear{{Applegate} et~al.,}{{Applegate}
  et~al.}{2014}]{Applegate2014}
{Applegate} D.~E.,  et~al., 2014, \mn@doi [\mnras] {10.1093/mnras/stt2129},
  \href {http://adsabs.harvard.edu/abs/2014MNRAS.439...48A} {439, 48}

\bibitem[\protect\citeauthoryear{{Bacon}, {Refregier}  \& {Ellis}}{{Bacon}
  et~al.}{2000}]{Bacon2000}
{Bacon} D.~J.,  {Refregier} A.~R.,   {Ellis} R.~S.,  2000, \mn@doi [MNRAS]
  {10.1046/j.1365-8711.2000.03851.x}, 318, 625

\bibitem[\protect\citeauthoryear{{Baker}, {Clampitt}, {Jain}  \&
  {Trodden}}{{Baker} et~al.}{2018}]{Baker2018}
{Baker} T.,  {Clampitt} J.,  {Jain} B.,   {Trodden} M.,  2018, \mn@doi [PRD]
  {10.1103/PhysRevD.98.023511}, \href
  {https://ui.adsabs.harvard.edu/\#abs/2018PhRvD..98b3511B} {98, 023511}

\bibitem[\protect\citeauthoryear{{Barreira}, {Cautun}, {Li}, {Baugh}  \&
  {Pascoli}}{{Barreira} et~al.}{2015}]{Barreira2015}
{Barreira} A.,  {Cautun} M.,  {Li} B.,  {Baugh} C.~M.,   {Pascoli} S.,  2015,
  \mn@doi [JCAP] {10.1088/1475-7516/2015/08/028}, \href
  {https://ui.adsabs.harvard.edu/abs/2015JCAP...08..028B} {2015, 028}

\bibitem[\protect\citeauthoryear{Barreira, Bose, Li  \& Llinares}{Barreira
  et~al.}{2017}]{Barreira2017}
Barreira A.,  Bose S.,  Li B.,   Llinares C.,  2017, \mn@doi [JCAP]
  {10.1088/1475-7516/2017/02/031}, 2017, 031

\bibitem[\protect\citeauthoryear{{Bartelmann} \& {Schneider}}{{Bartelmann} \&
  {Schneider}}{2001}]{Bartelmann2001}
{Bartelmann} M.,  {Schneider} P.,  2001, \mn@doi [\physrep]
  {10.1016/S0370-1573(00)00082-X}, \href
  {https://ui.adsabs.harvard.edu/\#abs/2001PhR...340..291B} {340, 291}

\bibitem[\protect\citeauthoryear{Bocquet et~al.}{Bocquet
  et~al.}{2018}]{Bocquet2018}
Bocquet S.,  et~al., 2018, preprint (\mn@eprint {arXiv} {1812.01679})

\bibitem[\protect\citeauthoryear{{Cai}, {Padilla}  \& {Li}}{{Cai}
  et~al.}{2015}]{Cai2015}
{Cai} Y.-C.,  {Padilla} N.,   {Li} B.,  2015, \mn@doi [\mnras]
  {10.1093/mnras/stv777}, \href
  {https://ui.adsabs.harvard.edu/abs/2015MNRAS.451.1036C} {451, 1036}

\bibitem[\protect\citeauthoryear{{Cardone}, {Camera}, {Mainini}, {Romano},
  {Diaferio}, {Maoli}  \& {Scaramella}}{{Cardone} et~al.}{2013}]{Cardone2013}
{Cardone} V.~F.,  {Camera} S.,  {Mainini} R.,  {Romano} A.,  {Diaferio} A.,
  {Maoli} R.,   {Scaramella} R.,  2013, \mn@doi [\mnras]
  {10.1093/mnras/stt084}, \href
  {https://ui.adsabs.harvard.edu/\#abs/2013MNRAS.430.2896C} {430, 2896}

\bibitem[\protect\citeauthoryear{Cautun, Li, Bose, Paillas, Armijo, Padilla  \&
  Cai}{Cautun et~al.}{2018}]{Cautun2018}
Cautun M.,  Li B.,  Bose S.,  Paillas E.,  Armijo J.,  Padilla N.,   Cai Y.-C.,
   2018, \mn@doi [\mnras] {10.1093/mnras/sty463}, 476, 3195

\bibitem[\protect\citeauthoryear{{Clampitt} \& {Jain}}{{Clampitt} \&
  {Jain}}{2015}]{Clampitt2015}
{Clampitt} J.,  {Jain} B.,  2015, \mn@doi [\mnras] {10.1093/mnras/stv2215},
  \href {https://ui.adsabs.harvard.edu/abs/2015MNRAS.454.3357C} {454, 3357}

\bibitem[\protect\citeauthoryear{{Clifton}, {Ferreira}, {Padilla}  \&
  {Skordis}}{{Clifton} et~al.}{2012}]{Clifton2012}
{Clifton} T.,  {Ferreira} P.~G.,  {Padilla} A.,   {Skordis} C.,  2012, \mn@doi
  [\physrep] {10.1016/j.physrep.2012.01.001}, \href
  {https://ui.adsabs.harvard.edu/abs/2012PhR...513....1C} {513, 1}

\bibitem[\protect\citeauthoryear{Coulton, Liu, Madhavacheril, Böhm  \&
  Spergel}{Coulton et~al.}{2018}]{Coulton2018}
Coulton W.~R.,  Liu J.,  Madhavacheril M.~S.,  Böhm V.,   Spergel D.~N.,
  2018, preprint (\mn@eprint {arXiv} {1810.02374})

\bibitem[\protect\citeauthoryear{{Davies}, {Cautun}  \& {Li}}{{Davies}
  et~al.}{2018}]{Davies2018}
{Davies} C.~T.,  {Cautun} M.,   {Li} B.,  2018, \mn@doi [\mnras]
  {10.1093/mnrasl/sly135}, \href
  {https://ui.adsabs.harvard.edu/\#abs/2018MNRAS.480L.101D} {480, L101}

\bibitem[\protect\citeauthoryear{{Davis} \& {Peebles}}{{Davis} \&
  {Peebles}}{1983}]{Davis1983}
{Davis} M.,  {Peebles} P.~J.~E.,  1983, \mn@doi [\apj] {10.1086/160884}, \href
  {http://adsabs.harvard.edu/abs/1983ApJ...267..465D} {267, 465}

\bibitem[\protect\citeauthoryear{Dietrich \& Hartlap}{Dietrich \&
  Hartlap}{2010}]{Dietrich2010}
Dietrich J.~P.,  Hartlap J.,  2010, \mn@doi [MNRAS]
  {10.1111/j.1365-2966.2009.15948.x}, 402, 1049

\bibitem[\protect\citeauthoryear{{Fan}, {Shan}  \& {Liu}}{{Fan}
  et~al.}{2010}]{Fan2010}
{Fan} Z.,  {Shan} H.,   {Liu} J.,  2010, \mn@doi [\apj]
  {10.1088/0004-637X/719/2/1408}, \href
  {https://ui.adsabs.harvard.edu/abs/2010ApJ...719.1408F} {719, 1408}

\bibitem[\protect\citeauthoryear{{Fu} et~al.,}{{Fu} et~al.}{2008}]{Fu2008}
{Fu} L.,  et~al., 2008, \mn@doi [\aap] {10.1051/0004-6361:20078522}, \href
  {https://ui.adsabs.harvard.edu/\#abs/2008A&A...479....9F} {479, 9}

\bibitem[\protect\citeauthoryear{{Giocoli}, {Moscardini}, {Baldi}, {Meneghetti}
   \& {Metcalf}}{{Giocoli} et~al.}{2018}]{Giocoli2018}
{Giocoli} C.,  {Moscardini} L.,  {Baldi} M.,  {Meneghetti} M.,   {Metcalf}
  R.~B.,  2018, \mn@doi [\mnras] {10.1093/mnras/sty1312}, \href
  {https://ui.adsabs.harvard.edu/\#abs/2018MNRAS.478.5436G} {478, 5436}

\bibitem[\protect\citeauthoryear{{Gruen} et~al.,}{{Gruen}
  et~al.}{2014}]{Gruen2014}
{Gruen} D.,  et~al., 2014, \mn@doi [\mnras] {10.1093/mnras/stu949}, \href
  {http://adsabs.harvard.edu/abs/2014MNRAS.442.1507G} {442, 1507}

\bibitem[\protect\citeauthoryear{{Gupta}, {Matilla}, {Hsu}  \&
  {Haiman}}{{Gupta} et~al.}{2018}]{Gupta2018}
{Gupta} A.,  {Matilla} J.~M.~Z.,  {Hsu} D.,   {Haiman} Z.,  2018, \mn@doi [PRD]
  {10.1103/PhysRevD.97.103515}, \href
  {http://adsabs.harvard.edu/abs/2018PhRvD..97j3515G} {97, 103515}

\bibitem[\protect\citeauthoryear{Hamana, Yoshida  \& Takada}{Hamana
  et~al.}{2004}]{Hamana2004}
Hamana T.,  Yoshida N.,   Takada M.,  2004, \mn@doi [MNRAS]
  {10.1111/j.1365-2966.2004.07691.x}, 350, 893

\bibitem[\protect\citeauthoryear{Hamana, Oguri, Shirasaki  \& Sato}{Hamana
  et~al.}{2012}]{Hamana2012}
Hamana T.,  Oguri M.,  Shirasaki M.,   Sato M.,  2012, \mn@doi [MNRAS]
  {10.1111/j.1365-2966.2012.21582.x}, 425, 2287

\bibitem[\protect\citeauthoryear{{Hamana}, {Sakurai}, {Koike}  \&
  {Miller}}{{Hamana} et~al.}{2015}]{Haman2015}
{Hamana} T.,  {Sakurai} J.,  {Koike} M.,   {Miller} L.,  2015, \mn@doi [\pasj]
  {10.1093/pasj/psv034}, \href {http://ads.nao.ac.jp/abs/2015PASJ...67...34H}
  {67, 34}

\bibitem[\protect\citeauthoryear{{Hamilton}}{{Hamilton}}{1993}]{Hamilton1993}
{Hamilton} A.~J.~S.,  1993, \mn@doi [\apj] {10.1086/173288}, \href
  {http://adsabs.harvard.edu/abs/1993ApJ...417...19H} {417, 19}

\bibitem[\protect\citeauthoryear{Hennawi \& Spergel}{Hennawi \&
  Spergel}{2005}]{Hennawi2005}
Hennawi J.~F.,  Spergel D.~N.,  2005, \mn@doi [ApJ] {10.1086/428749}, 624, 59

\bibitem[\protect\citeauthoryear{Heymans Catherine et~al.,}{Heymans
  et~al.}{2012}]{Heymans2012}
Heymans Catherine R.~B.,  et~al., 2012, \mn@doi [MNRAS]
  {10.1111/j.1365-2966.2012.21952.x}, 427, 146

\bibitem[\protect\citeauthoryear{{Higuchi} \& {Shirasaki}}{{Higuchi} \&
  {Shirasaki}}{2016}]{Higuchi2016}
{Higuchi} Y.,  {Shirasaki} M.,  2016, \mn@doi [\mnras] {10.1093/mnras/stw814},
  \href {https://ui.adsabs.harvard.edu/\#abs/2016MNRAS.459.2762H} {459, 2762}

\bibitem[\protect\citeauthoryear{{Hildebrandt} et~al.,}{{Hildebrandt}
  et~al.}{2017}]{Hildebrandt2017}
{Hildebrandt} H.,  et~al., 2017, \mn@doi [\mnras] {10.1093/mnras/stw2805},
  \href {http://adsabs.harvard.edu/abs/2017MNRAS.465.1454H} {465, 1454}

\bibitem[\protect\citeauthoryear{Hoekstra et~al.,}{Hoekstra
  et~al.}{2006}]{Hoekstra2006}
Hoekstra H.,  et~al., 2006, \mn@doi [\apj] {10.1086/503249}, 647, 116

\bibitem[\protect\citeauthoryear{{Hoekstra}, {Herbonnet}, {Muzzin}, {Babul},
  {Mahdavi}, {Viola}  \& {Cacciato}}{{Hoekstra} et~al.}{2015}]{Hoekstra2015}
{Hoekstra} H.,  {Herbonnet} R.,  {Muzzin} A.,  {Babul} A.,  {Mahdavi} A.,
  {Viola} M.,   {Cacciato} M.,  2015, \mn@doi [\mnras] {10.1093/mnras/stv275},
  \href {http://adsabs.harvard.edu/abs/2015MNRAS.449..685H} {449, 685}

\bibitem[\protect\citeauthoryear{{Huterer}}{{Huterer}}{2010}]{Huterer2010}
{Huterer} D.,  2010, \mn@doi [GRG] {10.1007/s10714-010-1051-z}, \href
  {https://ui.adsabs.harvard.edu/\#abs/2010GReGr..42.2177H} {42, 2177}

\bibitem[\protect\citeauthoryear{{Jain} \& {Van Waerbeke}}{{Jain} \& {Van
  Waerbeke}}{2000}]{Jain2000}
{Jain} B.,  {Van Waerbeke} L.,  2000, \mn@doi [\apjl] {10.1086/312480}, \href
  {https://ui.adsabs.harvard.edu/abs/2000ApJ...530L...1J} {530, L1}

\bibitem[\protect\citeauthoryear{Kaiser, Wilson  \& Luppino}{Kaiser
  et~al.}{2000}]{Kaiser2000}
Kaiser N.,  Wilson G.,   Luppino G.~A.,  2000, preprint (\mn@eprint {arXiv}
  {astro-ph/0003338})

\bibitem[\protect\citeauthoryear{{Kilbinger} et~al.,}{{Kilbinger}
  et~al.}{2013}]{Kilbinger2013}
{Kilbinger} M.,  et~al., 2013, \mn@doi [\mnras] {10.1093/mnras/stt041}, \href
  {https://ui.adsabs.harvard.edu/\#abs/2013MNRAS.430.2200K} {430, 2200}

\bibitem[\protect\citeauthoryear{Kratochvil, Haiman  \& May}{Kratochvil
  et~al.}{2010}]{Kratochvil2010}
Kratochvil J.~M.,  Haiman Z.,   May M.,  2010, \mn@doi [PRD]
  {10.1103/PhysRevD.81.043519}, 81, 043519

\bibitem[\protect\citeauthoryear{{LSST Dark Energy Science
  Collaboration}}{{LSST Dark Energy Science Collaboration}}{2012}]{LSST2012}
{LSST Dark Energy Science Collaboration} 2012, preprint, \href
  {https://ui.adsabs.harvard.edu/\#abs/2012arXiv1211.0310L} {} (\mn@eprint
  {arXiv} {1211.0310})

\bibitem[\protect\citeauthoryear{{LSST Science Collaboration} et~al.,}{{LSST
  Science Collaboration} et~al.}{2009}]{LSST2009}
{LSST Science Collaboration} et~al., 2009, preprint, \href
  {https://ui.adsabs.harvard.edu/\#abs/2009arXiv0912.0201L} {} (\mn@eprint
  {arXiv} {0912.0201})

\bibitem[\protect\citeauthoryear{{Landy} \& {Szalay}}{{Landy} \&
  {Szalay}}{1993}]{Landy1993}
{Landy} S.~D.,  {Szalay} A.~S.,  1993, \mn@doi [\apj] {10.1086/172900}, \href
  {http://adsabs.harvard.edu/abs/1993ApJ...412...64L} {412, 64}

\bibitem[\protect\citeauthoryear{{Li}, {Liu}, {Zorrilla Matilla}  \&
  {Coulton}}{{Li} et~al.}{2018}]{Li2018}
{Li} Z.,  {Liu} J.,  {Zorrilla Matilla} J.~M.,   {Coulton} W.~R.,  2018,
  preprint, \href {https://ui.adsabs.harvard.edu/\#abs/2018arXiv181001781L} {}
  (\mn@eprint {arXiv} {1810.01781})

\bibitem[\protect\citeauthoryear{{Lin} \& {Kilbinger}}{{Lin} \&
  {Kilbinger}}{2015}]{Lin2015}
{Lin} C.-A.,  {Kilbinger} M.,  2015, \mn@doi [\aap]
  {10.1051/0004-6361/201425188}, \href
  {https://ui.adsabs.harvard.edu/\#abs/2015A&A...576A..24L} {576, A24}

\bibitem[\protect\citeauthoryear{{Liu} \& {Haiman}}{{Liu} \&
  {Haiman}}{2016}]{J.Liu2016}
{Liu} J.,  {Haiman} Z.,  2016, \mn@doi [PRD] {10.1103/PhysRevD.94.043533},
  \href {https://ui.adsabs.harvard.edu/\#abs/2016PhRvD..94d3533L} {94, 043533}

\bibitem[\protect\citeauthoryear{{Liu}, {Petri}, {Haiman}, {Hui}, {Kratochvil}
  \& {May}}{{Liu} et~al.}{2015a}]{J.Liu2015}
{Liu} J.,  {Petri} A.,  {Haiman} Z.,  {Hui} L.,  {Kratochvil} J.~M.,   {May}
  M.,  2015a, \mn@doi [PRD] {10.1103/PhysRevD.91.063507}, \href
  {https://ui.adsabs.harvard.edu/\#abs/2015PhRvD..91f3507L} {91, 063507}

\bibitem[\protect\citeauthoryear{Liu et~al.,}{Liu et~al.}{2015b}]{X.Liu2015}
Liu X.,  et~al., 2015b, \mn@doi [\mnras] {10.1093/mnras/stv784}, \href
  {https://ui.adsabs.harvard.edu/\#abs/2015MNRAS.450.2888L} {450, 2888}

\bibitem[\protect\citeauthoryear{{Liu}, {Hill}, {Sherwin}, {Petri}, {B{\"o}hm}
  \& {Haiman}}{{Liu} et~al.}{2016a}]{J.Liu2016b}
{Liu} J.,  {Hill} J.~C.,  {Sherwin} B.~D.,  {Petri} A.,  {B{\"o}hm} V.,
  {Haiman} Z.,  2016a, \mn@doi [PRD] {10.1103/PhysRevD.94.103501}, \href
  {https://ui.adsabs.harvard.edu/abs/2016PhRvD..94j3501L} {94, 103501}

\bibitem[\protect\citeauthoryear{{Liu} et~al.,}{{Liu}
  et~al.}{2016b}]{X.Liu2016}
{Liu} X.,  et~al., 2016b, \mn@doi [PRL] {10.1103/PhysRevLett.117.051101}, \href
  {https://ui.adsabs.harvard.edu/\#abs/2016PhRvL.117e1101L} {117, 051101}

\bibitem[\protect\citeauthoryear{{Mantz} et~al.,}{{Mantz}
  et~al.}{2015}]{Mantz2015}
{Mantz} A.~B.,  et~al., 2015, \mn@doi [\mnras] {10.1093/mnras/stu2096}, \href
  {https://ui.adsabs.harvard.edu/abs/2015MNRAS.446.2205M} {446, 2205}

\bibitem[\protect\citeauthoryear{Marian, Schneider, Smith  \& Hilbert}{Marian
  et~al.}{2012}]{Marian2012}
Marian L.,  Schneider P.,  Smith R.~E.,   Hilbert S.,  2012, \mn@doi [MNRAS]
  {10.1111/j.1365-2966.2012.20992.x}, 423, 1711

\bibitem[\protect\citeauthoryear{{Marian}, {Smith}, {Hilbert}  \&
  {Schneider}}{{Marian} et~al.}{2013}]{Marian2013}
{Marian} L.,  {Smith} R.~E.,  {Hilbert} S.,   {Schneider} P.,  2013, \mn@doi
  [\mnras] {10.1093/mnras/stt552}, \href
  {https://ui.adsabs.harvard.edu/abs/2013MNRAS.432.1338M} {432, 1338}

\bibitem[\protect\citeauthoryear{{Marques}, {Liu}, {Zorrilla Matilla},
  {Haiman}, {Bernui}  \& {Novaes}}{{Marques} et~al.}{2018}]{Marques2018}
{Marques} G.~A.,  {Liu} J.,  {Zorrilla Matilla} J.~M.,  {Haiman} Z.,  {Bernui}
  A.,   {Novaes} C.~P.,  2018, preprint, \href
  {https://ui.adsabs.harvard.edu/\#abs/2018arXiv181208206M} {} (\mn@eprint
  {arXiv} {1812.08206})

\bibitem[\protect\citeauthoryear{{Maturi}, {Meneghetti}, {Bartelmann}, {Dolag}
  \& {Moscardini}}{{Maturi} et~al.}{2005}]{Maturi2005}
{Maturi} M.,  {Meneghetti} M.,  {Bartelmann} M.,  {Dolag} K.,   {Moscardini}
  L.,  2005, \mn@doi [\aap] {10.1051/0004-6361:20042600}, \href
  {https://ui.adsabs.harvard.edu/\#abs/2005A&A...442..851M} {442, 851}

\bibitem[\protect\citeauthoryear{{Melchior}, {Sutter}, {Sheldon}, {Krause}  \&
  {Wandelt}}{{Melchior} et~al.}{2014}]{Melchior2014}
{Melchior} P.,  {Sutter} P.~M.,  {Sheldon} E.~S.,  {Krause} E.,   {Wandelt}
  B.~D.,  2014, \mn@doi [\mnras] {10.1093/mnras/stu456}, \href
  {https://ui.adsabs.harvard.edu/abs/2014MNRAS.440.2922M} {440, 2922}

\bibitem[\protect\citeauthoryear{\"{O}ztireli \& Gross}{\"{O}ztireli \&
  Gross}{2012}]{Oztireli2012}
\"{O}ztireli A.~C.,  Gross M.,  2012, \mn@doi [ACM Trans. Graph.]
  {10.1145/2366145.2366189}, 31, 170:1

\bibitem[\protect\citeauthoryear{Paillas, Cautun, Li, Cai, Padilla, Armijo  \&
  Bose}{Paillas et~al.}{2019}]{Cautun2019}
Paillas E.,  Cautun M.,  Li B.,  Cai Y.-C.,  Padilla N.,  Armijo J.,   Bose S.,
   2019, \mn@doi [\mnras] {10.1093/mnras/stz022}, 484, 1149

\bibitem[\protect\citeauthoryear{{Peebles} \& {Hauser}}{{Peebles} \&
  {Hauser}}{1974}]{Peebles1974}
{Peebles} P.~J.~E.,  {Hauser} M.~G.,  1974, \mn@doi [\apjs] {10.1086/190308},
  \href {http://adsabs.harvard.edu/abs/1974ApJS...28...19P} {28, 19}

\bibitem[\protect\citeauthoryear{{Peel}, {Pettorino}, {Giocoli}, {Starck}  \&
  {Baldi}}{{Peel} et~al.}{2018}]{Peel2018}
{Peel} A.,  {Pettorino} V.,  {Giocoli} C.,  {Starck} J.-L.,   {Baldi} M.,
  2018, \mn@doi [\aap] {10.1051/0004-6361/201833481}, \href
  {https://ui.adsabs.harvard.edu/\#abs/2018A&A...619A..38P} {619, A38}

\bibitem[\protect\citeauthoryear{Pen, Zhang, van Waerbeke, Mellier, Zhang  \&
  Dubinski}{Pen et~al.}{2003}]{Pen2003}
Pen U.-L.,  Zhang T.,  van Waerbeke L.,  Mellier Y.,  Zhang P.,   Dubinski J.,
  2003, \mn@doi [ApJ] {10.1086/375734}, 592, 664

\bibitem[\protect\citeauthoryear{{Petri}}{{Petri}}{2016}]{Petri2016}
{Petri} A.,  2016, \mn@doi [A&C] {10.1016/j.ascom.2016.06.001}, \href
  {https://ui.adsabs.harvard.edu/\#abs/2016A&C....17...73P} {17, 73}

\bibitem[\protect\citeauthoryear{{Pizzuti} et~al.,}{{Pizzuti}
  et~al.}{2017}]{Pizzuti2017}
{Pizzuti} L.,  et~al., 2017, \mn@doi [JCAP] {10.1088/1475-7516/2017/07/023},
  \href {https://ui.adsabs.harvard.edu/abs/2017JCAP...07..023P} {2017, 023}

\bibitem[\protect\citeauthoryear{{Planck Collaboration} et~al.,}{{Planck
  Collaboration} et~al.}{2018}]{Planck2018}
{Planck Collaboration} et~al., 2018, preprint, \href
  {https://ui.adsabs.harvard.edu/\#abs/2018arXiv180706209P} {} (\mn@eprint
  {arXiv} {1807.06209})

\bibitem[\protect\citeauthoryear{{Refregier}, {Amara}, {Kitching}, {Rassat},
  {Scaramella}, {Weller}  \& {Euclid Imaging Consortium}}{{Refregier}
  et~al.}{2010}]{Refregier2010}
{Refregier} A.,  {Amara} A.,  {Kitching} T.~D.,  {Rassat} A.,  {Scaramella} R.,
   {Weller} J.,   {Euclid Imaging Consortium} f.~t.,  2010, preprint, \href
  {https://ui.adsabs.harvard.edu/\#abs/2010arXiv1001.0061R} {} (\mn@eprint
  {arXiv} {1001.0061})

\bibitem[\protect\citeauthoryear{{S{\'a}nchez} et~al.,}{{S{\'a}nchez}
  et~al.}{2017}]{Sanchez2017}
{S{\'a}nchez} C.,  et~al., 2017, \mn@doi [\mnras] {10.1093/mnras/stw2745},
  \href {https://ui.adsabs.harvard.edu/\#abs/2017MNRAS.465..746S} {465, 746}

\bibitem[\protect\citeauthoryear{{Schmidt}}{{Schmidt}}{2008}]{Schmidt2008}
{Schmidt} F.,  2008, \mn@doi [PRD] {10.1103/PhysRevD.78.043002}, \href
  {https://ui.adsabs.harvard.edu/\#abs/2008PhRvD..78d3002S} {78, 043002}

\bibitem[\protect\citeauthoryear{{Schneider}, {van Waerbeke}, {Kilbinger}  \&
  {Mellier}}{{Schneider} et~al.}{2002}]{Schneider2002}
{Schneider} P.,  {van Waerbeke} L.,  {Kilbinger} M.,   {Mellier} Y.,  2002,
  \mn@doi [\aap] {10.1051/0004-6361:20021341}, \href
  {https://ui.adsabs.harvard.edu/\#abs/2002A&A...396....1S} {396, 1}

\bibitem[\protect\citeauthoryear{{Semboloni} et~al.,}{{Semboloni}
  et~al.}{2006}]{Semboloni2006}
{Semboloni} E.,  et~al., 2006, \mn@doi [\aap] {10.1051/0004-6361:20054479},
  \href {https://ui.adsabs.harvard.edu/\#abs/2006A&A...452...51S} {452, 51}

\bibitem[\protect\citeauthoryear{Shan et~al.,}{Shan et~al.}{2012}]{Shan2012}
Shan H.,  et~al., 2012, \mn@doi [ApJ] {10.1088/0004-637x/748/1/56}, 748, 56

\bibitem[\protect\citeauthoryear{Shan et~al.,}{Shan et~al.}{2014}]{Shan2014}
Shan H.,  et~al., 2014, \mn@doi [MNRAS] {10.1093/mnras/stu1040}, 442, 2534

\bibitem[\protect\citeauthoryear{{Shan} et~al.,}{{Shan}
  et~al.}{2018}]{Shan2018}
{Shan} H.,  et~al., 2018, \mn@doi [\mnras] {10.1093/mnras/stx2837}, \href
  {https://ui.adsabs.harvard.edu/\#abs/2018MNRAS.474.1116S} {474, 1116}

\bibitem[\protect\citeauthoryear{{Shirasaki}}{{Shirasaki}}{2017}]{Shirasaki2017}
{Shirasaki} M.,  2017, \mn@doi [\mnras] {10.1093/mnras/stw2950}, \href
  {https://ui.adsabs.harvard.edu/\#abs/2017MNRAS.465.1974S} {465, 1974}

\bibitem[\protect\citeauthoryear{{Shirasaki}, {Hamana}  \&
  {Yoshida}}{{Shirasaki} et~al.}{2015}]{Shirasaki2015}
{Shirasaki} M.,  {Hamana} T.,   {Yoshida} N.,  2015, \mn@doi [\mnras]
  {10.1093/mnras/stv1854}, \href {http://ads.nao.ac.jp/abs/2015MNRAS.453.3043S}
  {453, 3043}

\bibitem[\protect\citeauthoryear{Shirasaki, Nishimichi, Li  \&
  Higuchi}{Shirasaki et~al.}{2017}]{Shirasaki:2016twn}
Shirasaki M.,  Nishimichi T.,  Li B.,   Higuchi Y.,  2017, \mn@doi [MNRAS]
  {10.1093/mnras/stw3254}, 466, 2402

\bibitem[\protect\citeauthoryear{{Shirasaki}, {Yoshida}  \&
  {Ikeda}}{{Shirasaki} et~al.}{2018}]{Shirasaki2018}
{Shirasaki} M.,  {Yoshida} N.,   {Ikeda} S.,  2018, preprint, \href
  {https://ui.adsabs.harvard.edu/\#abs/2018arXiv181205781S} {} (\mn@eprint
  {arXiv} {1812.05781})

\bibitem[\protect\citeauthoryear{{Takahashi}, {Hamana}, {Shirasaki},
  {Namikawa}, {Nishimichi}, {Osato}  \& {Shiroyama}}{{Takahashi}
  et~al.}{2017}]{Takahashi2017}
{Takahashi} R.,  {Hamana} T.,  {Shirasaki} M.,  {Namikawa} T.,  {Nishimichi}
  T.,  {Osato} K.,   {Shiroyama} K.,  2017, \mn@doi [\apj]
  {10.3847/1538-4357/aa943d}, \href
  {http://ads.nao.ac.jp/abs/2017ApJ...850...24T} {850, 24}

\bibitem[\protect\citeauthoryear{{Tsujikawa} \& {Tatekawa}}{{Tsujikawa} \&
  {Tatekawa}}{2008}]{Tsujikawa2008}
{Tsujikawa} S.,  {Tatekawa} T.,  2008, \mn@doi [PLB]
  {10.1016/j.physletb.2008.06.052}, \href
  {https://ui.adsabs.harvard.edu/\#abs/2008PhLB..665..325T} {665, 325}

\bibitem[\protect\citeauthoryear{{Tudorica} et~al.,}{{Tudorica}
  et~al.}{2017}]{Tudorica2017}
{Tudorica} A.,  et~al., 2017, \mn@doi [\aap] {10.1051/0004-6361/201731267},
  \href {https://ui.adsabs.harvard.edu/\#abs/2017A&A...608A.141T} {608, A141}

\bibitem[\protect\citeauthoryear{{Umetsu} et~al.,}{{Umetsu}
  et~al.}{2014}]{Umetsu2014}
{Umetsu} K.,  et~al., 2014, \mn@doi [\apj] {10.1088/0004-637X/795/2/163}, \href
  {http://adsabs.harvard.edu/abs/2014ApJ...795..163U} {795, 163}

\bibitem[\protect\citeauthoryear{{Umetsu}, {Zitrin}, {Gruen}, {Merten},
  {Donahue}  \& {Postman}}{{Umetsu} et~al.}{2016}]{Umetsu2016}
{Umetsu} K.,  {Zitrin} A.,  {Gruen} D.,  {Merten} J.,  {Donahue} M.,
  {Postman} M.,  2016, \mn@doi [\apj] {10.3847/0004-637X/821/2/116}, \href
  {https://ui.adsabs.harvard.edu/abs/2016ApJ...821..116U} {821, 116}

\bibitem[\protect\citeauthoryear{{Van Uitert} et~al.,}{{Van Uitert}
  et~al.}{2018}]{vanUitert2018}
{Van Uitert} E.,  et~al., 2018, \mn@doi [\mnras] {10.1093/mnras/sty551}, \href
  {https://ui.adsabs.harvard.edu/abs/2018MNRAS.476.4662V} {476, 4662}

\bibitem[\protect\citeauthoryear{{Van Waerbeke} et~al.,}{{Van Waerbeke}
  et~al.}{2000}]{VanWaerbeke2000}
{Van Waerbeke} L.,  et~al., 2000, \aap, \href
  {https://ui.adsabs.harvard.edu/abs/2000A&A...358...30V} {358, 30}

\bibitem[\protect\citeauthoryear{Van~Waerbeke et~al.,}{Van~Waerbeke
  et~al.}{2013}]{VanWaerbeke2013}
Van~Waerbeke L.,  et~al., 2013, \mn@doi [MNRAS] {10.1093/mnras/stt971}, 433,
  3373

\bibitem[\protect\citeauthoryear{{Von der Linden} et~al.,}{{Von der Linden}
  et~al.}{2014}]{vonderLinden2014}
{Von der Linden} A.,  et~al., 2014, \mn@doi [\mnras] {10.1093/mnras/stt1945},
  \href {http://adsabs.harvard.edu/abs/2014MNRAS.439....2V} {439, 2}

\bibitem[\protect\citeauthoryear{{Wei}, {Li}, {Kang}, {Liu}, {Fan}, {Yuan}  \&
  {Pan}}{{Wei} et~al.}{2018}]{Wei2018}
{Wei} C.,  {Li} G.,  {Kang} X.,  {Liu} X.,  {Fan} Z.,  {Yuan} S.,   {Pan} C.,
  2018, \mn@doi [\mnras] {10.1093/mnras/sty1268}, \href
  {https://ui.adsabs.harvard.edu/\#abs/2018MNRAS.478.2987W} {478, 2987}

\bibitem[\protect\citeauthoryear{{Weinberg}, {Mortonson}, {Eisenstein},
  {Hirata}, {Riess}  \& {Rozo}}{{Weinberg} et~al.}{2013}]{Weinberg2013}
{Weinberg} D.~H.,  {Mortonson} M.~J.,  {Eisenstein} D.~J.,  {Hirata} C.,
  {Riess} A.~G.,   {Rozo} E.,  2013, \mn@doi [\physrep]
  {10.1016/j.physrep.2013.05.001}, \href
  {http://adsabs.harvard.edu/abs/2013PhR...530...87W} {530, 87}

\bibitem[\protect\citeauthoryear{{Wittman}, {Tyson}, {Kirkman}, {Dell'Antonio}
  \& {Bernstein}}{{Wittman} et~al.}{2000}]{Wittman2000}
{Wittman} D.~M.,  {Tyson} J.~A.,  {Kirkman} D.,  {Dell'Antonio} I.,
  {Bernstein} G.,  2000, \mn@doi [\nat] {10.1038/35012001}, \href
  {http://adsabs.harvard.edu/abs/2000Natur.405..143W} {405, 143}

\bibitem[\protect\citeauthoryear{{Yang}, {Kratochvil}, {Wang}, {Lim}, {Haiman}
  \& {May}}{{Yang} et~al.}{2011}]{Yang2011}
{Yang} X.,  {Kratochvil} J.~M.,  {Wang} S.,  {Lim} E.~A.,  {Haiman} Z.,   {May}
  M.,  2011, \mn@doi [PRD] {10.1103/PhysRevD.84.043529}, \href
  {https://ui.adsabs.harvard.edu/\#abs/2011PhRvD..84d3529Y} {84, 043529}

\bibitem[\protect\citeauthoryear{{Zorrilla Matilla}, {Haiman}, {Hsu}, {Gupta}
  \& {Petri}}{{Zorrilla Matilla} et~al.}{2016}]{Matilla2017}
{Zorrilla Matilla} J.~M.,  {Haiman} Z.,  {Hsu} D.,  {Gupta} A.,   {Petri} A.,
  2016, \mn@doi [PRD] {10.1103/PhysRevD.94.083506}, \href
  {http://adsabs.harvard.edu/abs/2016PhRvD..94h3506Z} {94, 083506}

\makeatother
\end{thebibliography}



\appendix

\section{Partitioning an all-sky map into smaller non-overlapping maps}\label{Appendix: split all sky map}

\begin{figure*}
    \centering
        \includegraphics[width=\columnwidth]{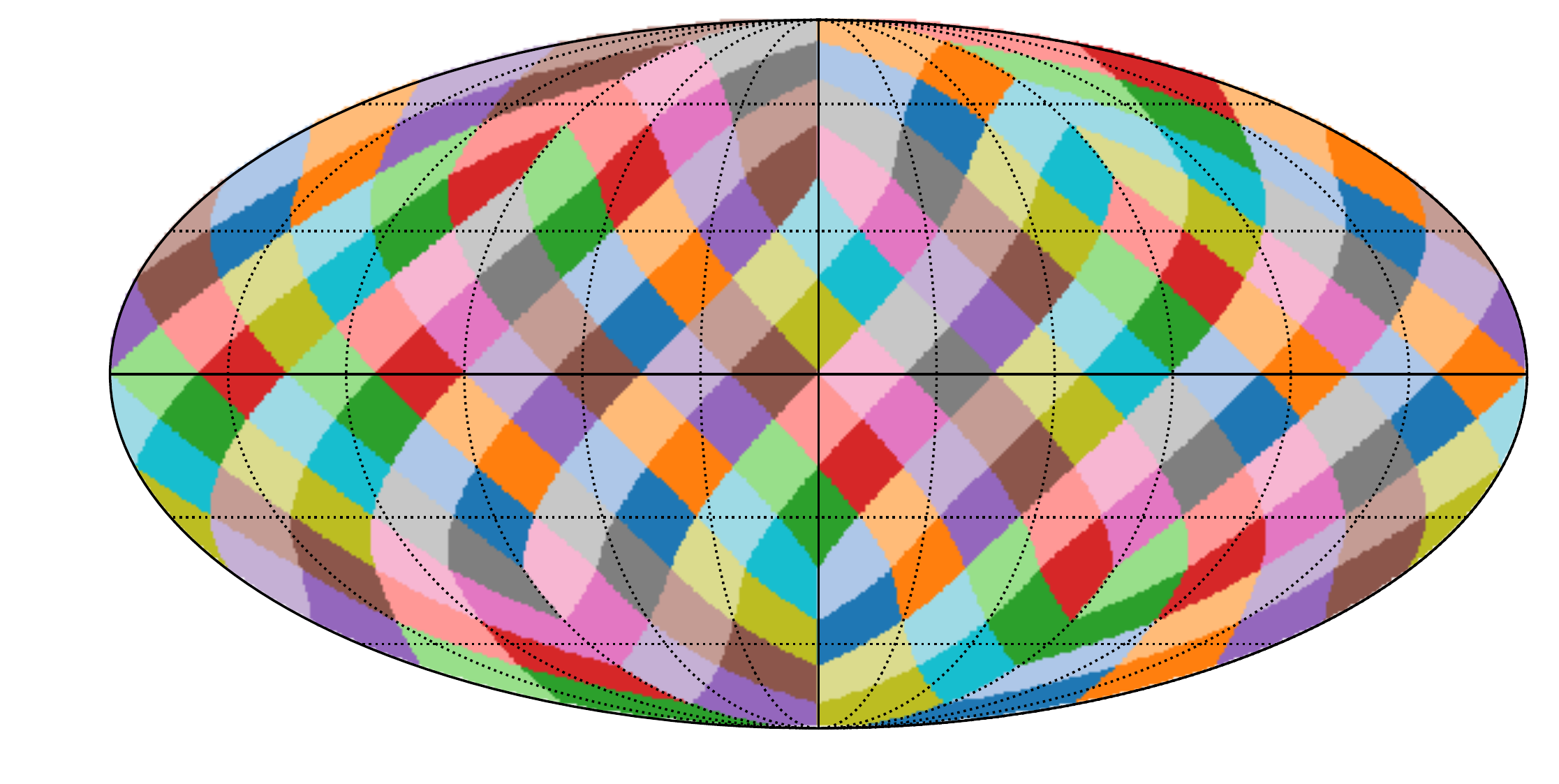}
        \includegraphics[width=\columnwidth]{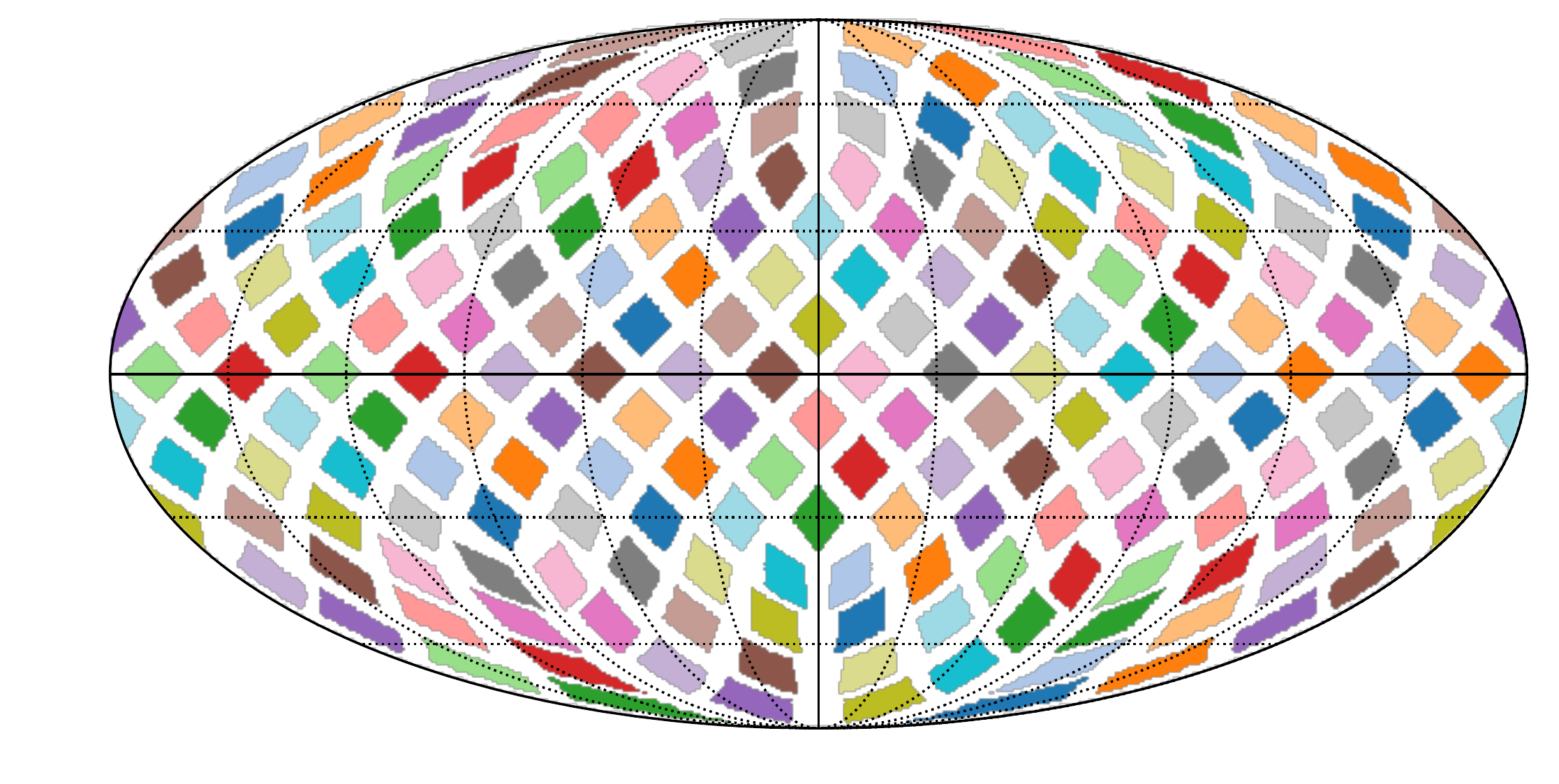}
    \caption{An illustration of our procedure of partitioning an all-sky map into smaller non-overlapping maps. We first tile the sky using a HEALPix grid with $N_{\rm{side}} = 4$. This step is shown in the left panel, with each coloured patch corresponding to a HEALPix pixel. Then, we further extract a \map{10} map from the centre of each HEALPix pixel. The resulting square maps are shown as coloured patches in the right panel. The  white space between the patches shows that our small maps are non-overlapping. 
    Each small square patch is then projected on to a plane tangential to their centre, giving us a \map{10} flat map.
    }
    \label{fig:splitsky}
\end{figure*}

The \citetalias{Takahashi2017} maps are all-sky maps with a HEALPix data structure, in which pixels are stored on the surface of a sphere. To simplify our analysis, we used the flat-sky approximation and thus we needed to partition the all-sky map into smaller, and preferably, non-overlapping maps. 

To achieve this, we capitalise on the HEALPix data structure and first define a set of coarser HEALPix pixels with a resolution of $N_{\rm{side}} = 4$, which corresponds to a pixel area of roughly $215\degSq$. We then assign each of the (higher resolution) data pixels to the coarser pixel that they are enclosed by. This is shown by the illustration on the left in Fig. \ref{fig:splitsky} (using Mollweide projection), where each coloured patch shows a course pixel. Next, for each sub region defined by the coarse pixels, we define a (flat) plane tangential to the centre of the coarse pixels and project the data pixels onto that plane. We then extract a square of \map{10} (centred on the centre of the plane) from each plane giving us 184 \map{10} flat maps. The HEALPix pixels that are projected onto the flat maps are converted into regular square pixels, where we interpolate between HEALPix pixels for square pixels that overlap with multiple HEALPix pixels. The benefit of this approach is that there is no overlap between any two maps as illustrated in the right panel of \ref{fig:splitsky}.

We note that a HEALPix resolution of $N_{\rm{side}} = 4$ actually gives 192 pixels, however due to the irregular shapes of HEALPix pixels (which arises from the requirement that all pixels have the same area), we find that 8 pixels have to be discarded since they cannot enclose squares of size \map{10}.

\section{Biased 2PCF estimation for small maps} 
\label{Appendix: small map bias}

Estimation of 2PCFs is straightforward in idealised situations. The 2PCF, $\xi(r)$, characterises the excess probability of finding a pair of tracers in two volume elements, ${\rm d}V_i$ and ${\rm d}V_j$, that are separated by a distance $r$: 
\begin{equation}
    {\rm d}P_{ij}(r) = \bar{n}^2\left[1+\xi(r)\right]{\rm d}V_i{\rm d}V_j,
\end{equation}
where $\bar{n}$ represents the expected tracer number density. In N-body simulations with periodic boundary conditions, as an example, $\bar{n}$ is the known mean number density and so the excess probability ${\rm d}P_{ij}$ can be evaluated by counting the number of pairs that are separated by a distance $r-\Delta r$ to $r+\Delta r$ and comparing that against $\bar{n}^2$. In realistic situations, $\bar{n}$ is not always known -- this can for example be due to the geometry, mask, fibre collision and redshift failure in a galaxy redshift survey, or the small map size with boundaries in our WL peak catalogues. The uncertainty in the expected number of tracers in a given volume can cause biased 2PCF estimations. It is known that, for examples, the \citet{Peebles1974} estimator
\begin{equation}
    \xi_{\rm PH}(\theta) = \bigg(\frac{N_R}{N_D}\bigg)^2 \frac{DD(\theta)}{RR(\theta)} - 1,
\end{equation}
and the \citet{Davis1983} estimator 
\begin{equation}
    \xi_{\rm DP}(\theta) = 2\frac{N_R}{N_D}\frac{DD(\theta)}{DR(\theta)} - 1,
\end{equation}
have errors that depend to the first order on the uncertainty of the expected tracer number density. On the other hand, the \citet{Hamilton1993} and the \citet{Landy1993} estimators have errors which are second order in this uncertainty and are more commonly used. In the above $N_D$ and $N_R$ are the numbers of data and random points, and $DD$, $DR$ and $RR$ are the numbers of data-data, data-random and random-random pairs in bins $\theta \pm \delta \theta$ respectively.

The Landy-Szalay estimator is given by, 
\begin{equation}
    \xi_{\rm{LS}}(\theta) = 1 + \bigg(\frac{N_R}{N_D}\bigg)^2 \frac{DD(\theta)}{RR(\theta)} - \bigg(\frac{N_R}{N_D}\bigg) \frac{DR(\theta)}{RR(\theta)}.
    \label{Appendix Eq: LS estimator}
\end{equation}
When analysing $n$ maps, there are $n$ different $N_D$, $DD$ and $DR$ values, that is one per map ($N_R$ and $RR$ can be taken as constants since the same random catalogue can be used for each map). We checked that our results are stable to a change in the number of randoms used. 

Given the expression of Eq.~\eqref{Appendix Eq: LS estimator}, there are two possible ways to calculate the mean 2PCF, $\langle\xi\rangle$, where $\langle\cdot\rangle$ denotes the mean value over the $n$ maps, given respectively by 
\begin{equation}
    \langle\xi_{\rm{LS}}(\theta) \rangle_1 = \left\langle 1 + \left(\frac{N_R}{N_D}\right)^2 \frac{DD}{RR} - \left(\frac{N_R}{N_D}\right) \frac{DR}{RR} \right\rangle,
    \label{Appendix Eq: LS estimator mean1}
\end{equation}
and
\begin{equation}
    \langle\xi_{\rm{LS}}(\theta)\rangle_2 = 1 + \left(\frac{\langle N_R\rangle}{\langle N_D\rangle}\right)^2 \frac{\langle DD\rangle}{\langle RR\rangle} - \left(\frac{\langle N_R\rangle}{\langle N_D\rangle}\right) \frac{\langle DR\rangle}{\langle RR\rangle},
    \label{Appendix Eq: LS estimator mean2}
\end{equation}
where we have dropped the $\theta$ dependence of $DD,DR$ and $RR$ to lighten the notations. 
For Eq.~\eqref{Appendix Eq: LS estimator mean1}, we calculate $n$ 2PCFs from the $n$ maps and take the mean value. For Eq.~\eqref{Appendix Eq: LS estimator mean2}, we first calculate the mean over all maps of $N_D$, $DD$ and $DR$, and then use these mean values to calculate the mean 2PCF. In general, these two approaches do not give identical results, that is 
\begin{equation}
     \langle\xi_{\rm{LS}}(\theta)\rangle_1 \neq  \langle\xi_{\rm{LS}}(\theta)\rangle_2.
\end{equation}
Naively, it seems natural to calculate the mean 2PCF using Eq.~\eqref{Appendix Eq: LS estimator mean1} -- after all, if we only had a single map, we would use this formula (excluding the outer $\langle\cdot\rangle$) to estimate the 2PCF. However, we found that this approach actually leads to biased estimates when the number of tracers varies considerably between the different maps. This is particularly the case when the number of peaks in a catalogue is low, and the effect is particularly strong for peaks with high $\nu$ values, for which the number density is low, and for small maps, such as the $\map{3.5}$ ones. 

\begin{figure}
    \centering
    \includegraphics[width=\columnwidth]{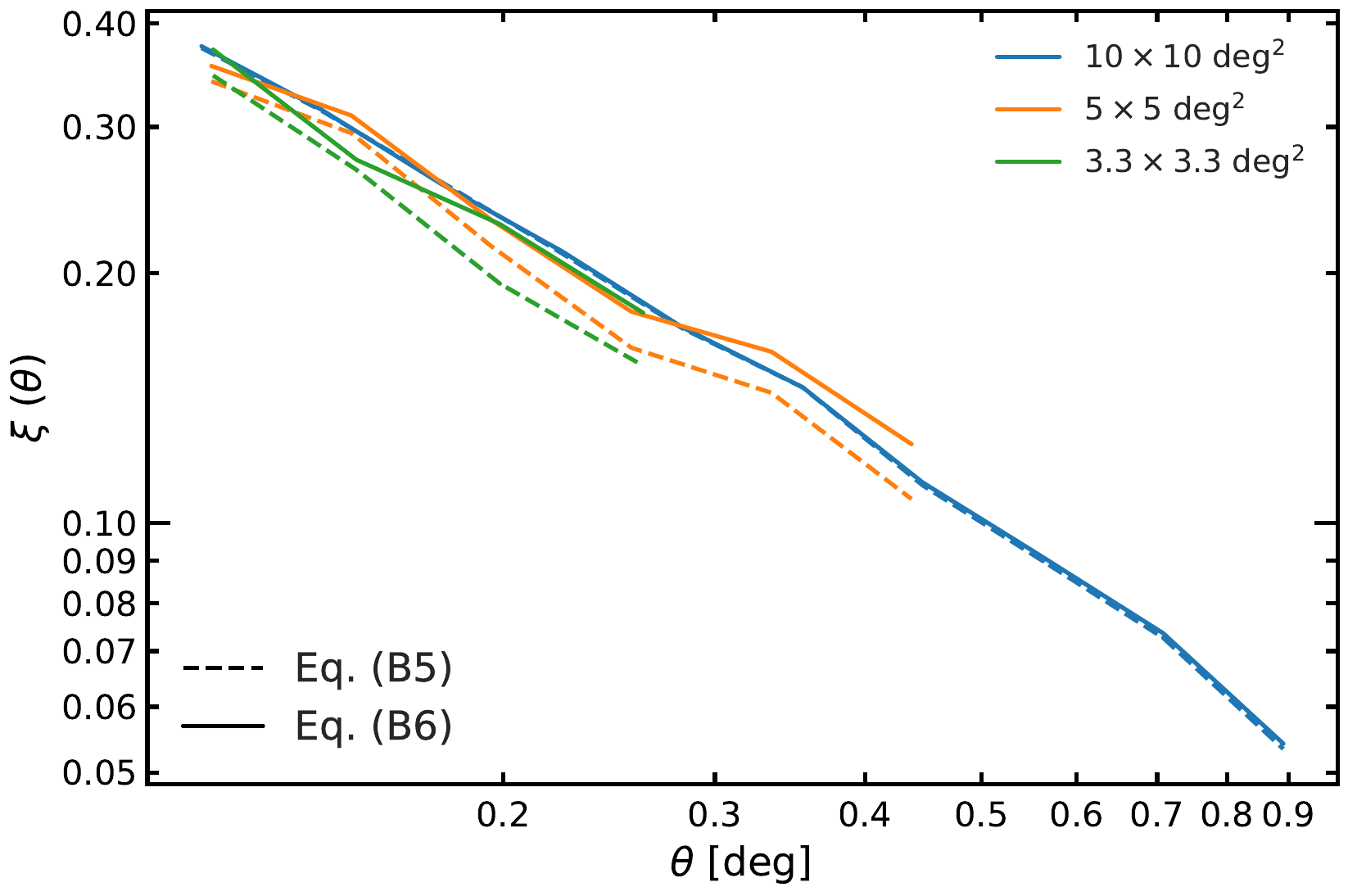}
    \vskip -.2cm
    \caption{Mean 2PCFs calculated using Eqs. \eqref{Appendix Eq: LS estimator mean1} (dashed) and \eqref{Appendix Eq: LS estimator mean2} (solid), for maps of size \map{10} (blue), \map{5} (orange) and \map{3.3} (green).
    }
    \label{fig:Appendix: mean tests}
\end{figure}

To see this point, we compare these two approaches as follows. First, we split each of the 184 \citetalias{Takahashi2017} maps, whose size is $\map{10}$, into four $\map{5}$ maps and nine $\map{3.3}$ maps, which give us in total 184, 736 and 1656 maps of the three sizes respectively. Then, using respectively Eqs.~\eqref{Appendix Eq: LS estimator mean1} and \eqref{Appendix Eq: LS estimator mean2}, we calculate the mean 2PCF for the three different maps sizes, and the results are shown in Fig.~\ref{fig:Appendix: mean tests}. 

Fig.~\ref{fig:Appendix: mean tests} clearly shows that, as the \citetalias{Takahashi2017} maps are split into progressively smaller sections, the mean 2PCF calculated using Eq.~\eqref{Appendix Eq: LS estimator mean1} drops in amplitude, whereas using Eq.~\eqref{Appendix Eq: LS estimator mean2} leads to a constant amplitude. The difference between the two approaches is small for the $\map{10}$ maps and only becomes significant for the smaller maps. This implies that the bias from Eq. \eqref{Appendix Eq: LS estimator mean1} depends on the map size, or more exactly the number of tracers used for the 2PCF estimation. We have performed similar tests for 3D galaxy 2PCFs and found a similar bias effect when using small box sizes. Finally, the mean 2PCFs from Eq. \eqref{Appendix Eq: LS estimator mean2} for the different maps sizes do not line up exactly, which is due to some pairs being lost at the sub-map boundaries as large maps are split up into smaller maps. We checked for this and found that the inclusion of cross sub-map pairs restores the original 2PCFs. 

Physically, the reason why Eq.~\eqref{Appendix Eq: LS estimator mean1} leads to biased 2PCF estimations is that the number of WL peaks per map is small and this translates into a large uncertainty in the mean tracer number density when estimated individually for each map. Even though this uncertainty enters the 2PCF estimation only at second order for the Landy-Szalay estimator, it can still strongly affect the latter. In contrast, Eq.~\eqref{Appendix Eq: LS estimator mean2} essentially treats the $n$ maps as a single (combined) one, for which the uncertainty in the expected mean peak number is small.

The biased 2PCF estimation using the Landy-Szalay estimator caused by the small tracer number is important for this study, since the WL maps from the \citetalias{Matilla2017} simulations have a map size of $\map{3.5}$, which is in the regime where the biasing effect is strong. As a result, in this paper we calculate the mean 2PCF using Eq.~\eqref{Appendix Eq: LS estimator mean2}.

\section{Error estimates} 
\label{Appendix: error estimation}

\begin{figure}
    \centering
    \includegraphics[width=\columnwidth]{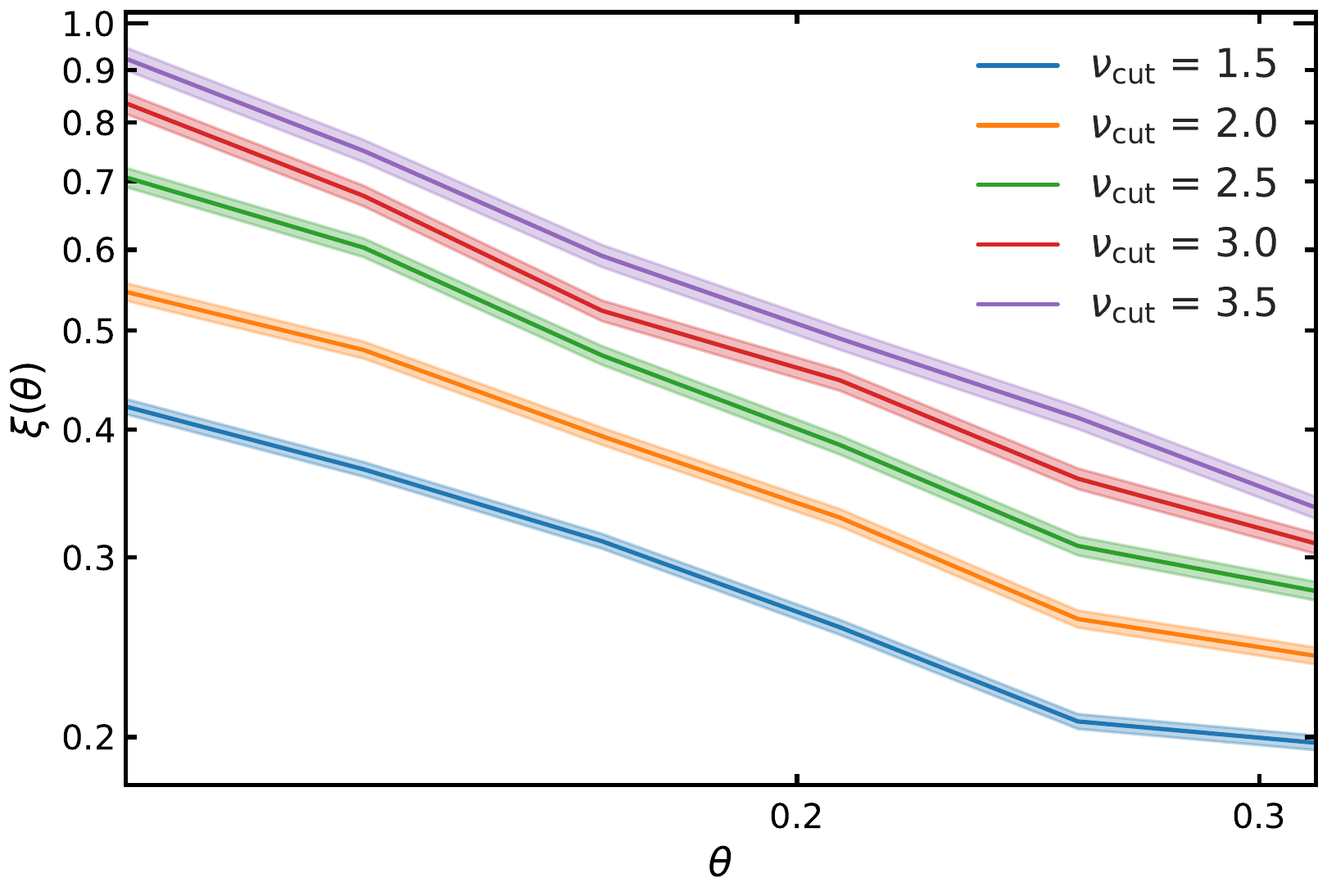}
    \vskip -.2cm
    \caption{The mean 2PCFs of 512 \map{3.5} maps extracted from the \citetalias{Takahashi2017} all sky map for peak catalogues with $\nu_{\rm{cut}}$ $\in$ [1.5,3.5]. The shaded regions show the jackknifed error bars.
    }
    \label{Appendix: fig: relative errors}
\end{figure}

For each of the \citetalias{Matilla2017} cosmologies we used the $N = $ 512 $3.5\times3.5$ deg$^2$ maps to evaluate the mean 2PCF $\langle\xi\rangle$. We estimate the standard error for $\langle\xi\rangle$ using the jackknife method, by calculating $N-1$ mean values from sequentially removing individual maps from the sample, and taking the standard deviation of the $N-1$ means from the 512 maps. However, we found the error to be significantly smaller than expected, of roughly $<1\%$ of $\langle\xi\rangle$ itself. On the other hand, when repeating the same practice on 512 $3.5\times3.5$ deg$^2$ maps extracted from the \citetalias{Takahashi2017} all-sky map, we found the standard errors were larger and more reasonable, of roughly $2$-$3\%$ of the mean 2PCF. 

This discrepancy in the magnitudes of the standard errors in the two different suites of maps is likely caused by the way in which the multiple convergence maps were generated. In the \citetalias{Matilla2017} case, the 512 maps were generated from multiple lines of sight by shifting, reorienting and tiling a single simulation box of size $240h^{-1}$Mpc, which means that the scatter in the different maps is likely to only contain an error component representing the line-of-sight variation. In contrast, the \citetalias{Takahashi2017} maps were all-sky maps generated using much larger boxes with minimal repetition of structures along the lines of sight, which means that these maps better sample the variation due to large-scale modes. The additional source of variance in the \citetalias{Takahashi2017} maps can explain the increase in their measured standard error. 

In order to have a more realistic estimate of the standard error associated to the \citetalias{Matilla2017} maps, we extract 512 $\map{3.5}$ maps from the \citetalias{Takahashi2017} all-sky map and use jackknife to find the error of the mean 2PCFs, $\langle\xi\rangle$. For illustration purposes, the resulting $\langle\xi\rangle$ and their errors for a few values of $\nucut$ are shown in Fig.~\ref{Appendix: fig: relative errors} as respectively lines and shaded regions.

We then take this relative error as our estimate of the standard error for the mean 2PCFs from the \citetalias{Matilla2017} maps, as a way to (approximately) include the contributions to the error from large-scale modes. 

The above estimate of the error associated to the \citetalias{Matilla2017} maps is likely to be an underestimate since the estimated error corresponds to the case when each of the 512 \citetalias{Matilla2017} maps would have been obtained from a different N-body simulation. However, this is not the case since all the \citetalias{Matilla2017} maps were obtained from the same simulation.
Thus, the errors used in this paper serve only as a way to gain rough indications of the quality of our models for the WL peak statistics, which we present as a proof of concept. In a future work, we plan to run a suite of large simulations similar to those used by \citetalias{Takahashi2017}, for different cosmological models, to further study the self-similar properties of the rescaled peak 2PCFs.

\section{The independence of self-similarity on the $\nu$ definition}

\begin{figure*}
    \centering
    \includegraphics[width = 2\columnwidth]{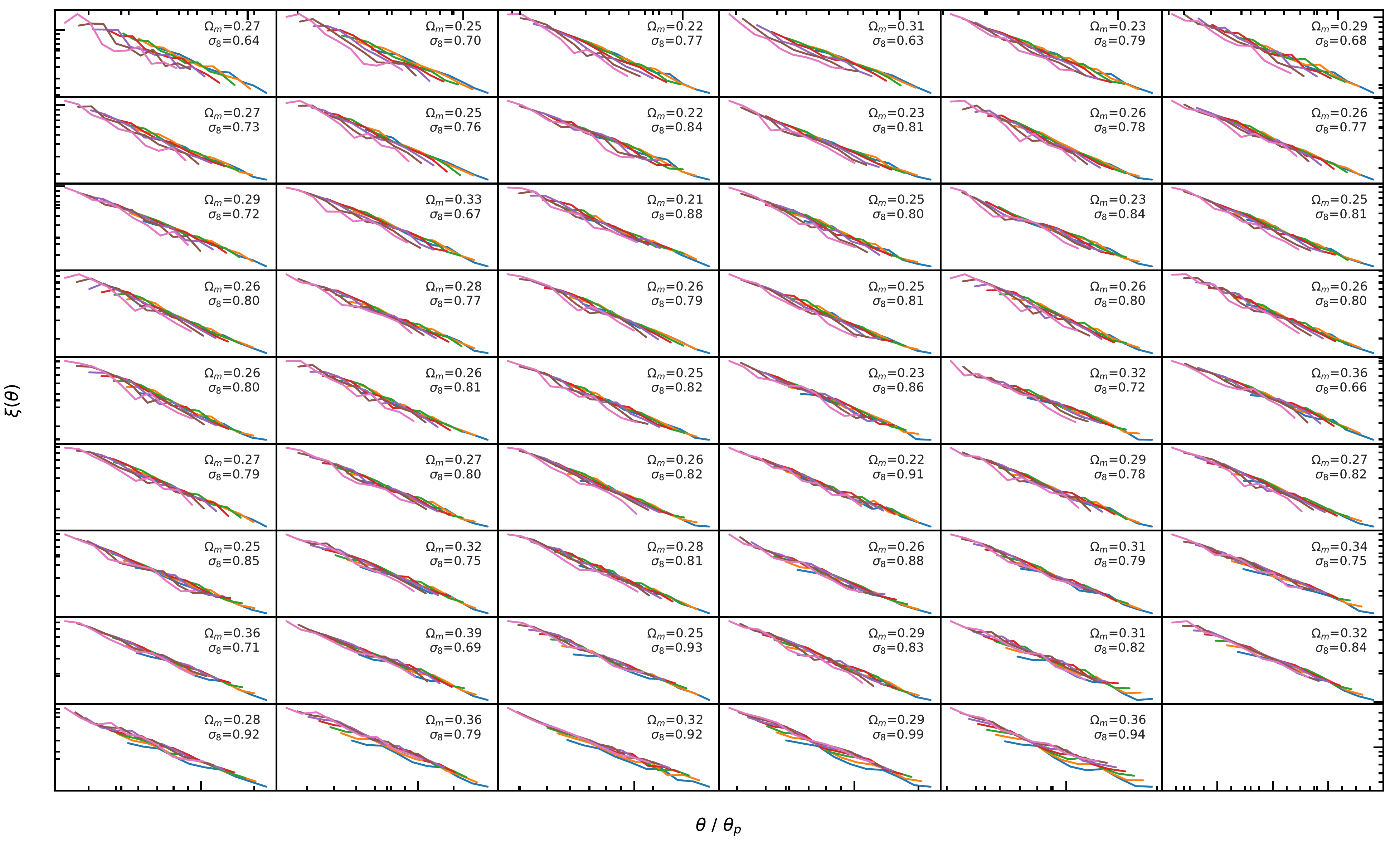}
    \vskip -.2cm
    \caption{The same as Fig.~\ref{fig:matrix plot}, except only the cosmological models shown as orange points in Fig.~\ref{fig:cosmo models} are plotted and the definition for $\nu$ is changed from Eq.~\eqref{Eq:nu} to $\nu = (\kappa - \mu) / \sigma_{\rm{GSN}}$ where $\sigma_{\rm{GSN}} = 0.013$ for all models.}
    \label{Appendix: fig: matrix plot const std}
\end{figure*}

In Eq.~\eqref{Eq:nu} we choose to define the SNR, $\nu$, in terms of a cosmology-dependent rms convergence, $\sigma$, which is analytically parameterised through a simple dependence on $(\Omega_m,\sigma_8)$, as exemplified in Fig.~\ref{fig:std cosmo}. Besides having a readily-predictable $\sigma$, this approach has the added benefit of allowing us to more naturally define the amplitude of WL peaks for a given cosmology relative to its own convergence rms, bearing in mind that the wide coverage of cosmological parameters means that the $\sigma$ values can vary by a factor of a few across the \citetalias{Matilla2017} maps; cf.~Fig.~\ref{fig:std cosmo}.

One can argue that given an observational WL map, the value of $\sigma$ receives contributions from both the physical convergence rms and the GSN, and that the actual value of $\sigma$ as measured from such noisy maps is a natural choice that can be used to define $\nu$. Such is the logic followed in Section \ref{section:GSN} where we analysed the rescaled peak 2PCFs in the GSN-added maps. Alternatively, one may argue that in real observations we do not necessarily know the true cosmology, but we do understand the survey specifications well enough to know the expected noise level. This leads to another natural way to define $\nu$, namely by using $\sigma\equiv\sigma_{\rm GSN}$. Given this flexibility in $\nu$ definition, we would like to check that the self-similar behaviour of the 2PCFs for the resulting peak catalogues is not affected by it. This is done in Fig.~\ref{Appendix: fig: matrix plot const std}, which is similar to Fig.~\ref{fig:matrix plot} but for a subset of cosmological models (the ones represented by the orange symbols in Fig.~\ref{fig:cosmo models}), and where $\nu$ is defined as $\nu = (\kappa - \mu) / \sigma_{\rm{GSN}}$ with $\sigma_{\rm{GSN}} = 0.013$, which corresponds to the rms of a GSN only map smoothed with $\theta_s = 1$ arcmin, is used for all models.

We find that the cosmology-dependent description of WL peak amplitude results in a (marginal) improvement of the self similarity of the 2PCFs for all cosmologies (shown in Fig.~\ref{fig:matrix plot}) compared to using a $\sigma$ definition that is constant across all cosmologies, which is shown in Fig.~\ref{Appendix: fig: matrix plot const std}. It can be seen that the self similarity of the 2PCFs worsens notably for some of the panels, which correspond to models with more extreme $(\Omega_m,\sigma_8)$ values. This is not surprising because, as mentioned above, the models studied here vary wildly in their $\sigma$ values, and by using a constant $\sigma_{\rm GSN}$ to define $\nu$ one is essentially selecting very different peak populations in them -- in the more extreme models the peaks that end up being selected do not possess the self-similarity (remember that this property is only present for a limited range of peak heights). Hence, by using the cosmology dependent form of $\nu$ in Eq.~\eqref{Eq:nu}, the 2PCF is self similar for a larger range of cosmologies. However, if one focuses on the more realistic $(\Omega_m,\sigma_8)$ parameters, then using a constant $\sigma_{\rm GSN}$ to define $\nu$ should not affect the potential of the rescaled peak 2PCF as a cosmological probe.


\bsp	
\label{lastpage}
\end{document}